\documentclass[aps, pre, twocolumn, groupedaddress]{revtex4-2}

\usepackage{amsfonts,amssymb,amsmath,bm,natbib}
\usepackage{graphicx,units}
\usepackage[hidelinks]{hyperref}

\DeclareMathOperator{\Tr}{Tr}

\begin{document}
	
	\title{Effects of liquid fraction and contact angle on structure and coarsening in two-dimensional foams}
	\author{Jacob Morgan}
	\email{jam164@aber.ac.uk}
	\author{Simon Cox}
	\affiliation{Department of Mathematics, Aberystwyth University, Penglais, Aberystwyth SY23 3BZ, UK}
	\date{December 13, 2023}

	\begin{abstract}
		Aqueous foams coarsen with time due to gas diffusion through the liquid. The mean bubble size grows, and small bubbles vanish. However, coarsening is little understood for foams with an intermediate liquid content, particularly in the presence of surfactant-induced attractive forces between the bubbles, measured by the contact angle. Rigorous bubble growth laws have yet to be developed, and the evolution of bulk foam properties is unclear. We present a quasi-static numerical model for coarsening in two-dimensional wet foams, focusing on growth laws and related bubble properties. The deformation of bubbles is modelled using a finite-element approach, and the gas flow through both films and Plateau borders is approximated. We give results for disordered two-dimensional wet foams with $256$ to $1024$ bubbles, at liquid fractions from~$2\%$ to beyond the zero-contact-angle jamming transition, and with contact angles up to $10^\circ$. Simple analytical models are developed to aid interpretation. We find that nonzero contact angle causes a proxy of the initial coarsening rate to plateau at large liquid fractions, and that the individual bubble growth rates are closely related to their effective number of neighbours.
	\end{abstract}

	\maketitle

	\section{Introduction \label{sec:intro}}
	
\begin{figure}
		\includegraphics[width=1.0\columnwidth]{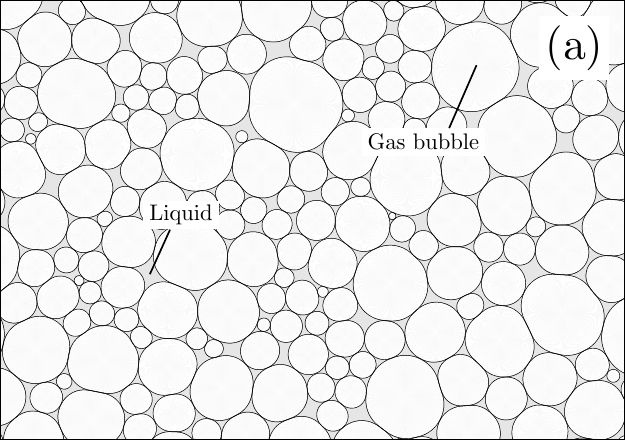}
		\includegraphics[width=1.0\columnwidth]{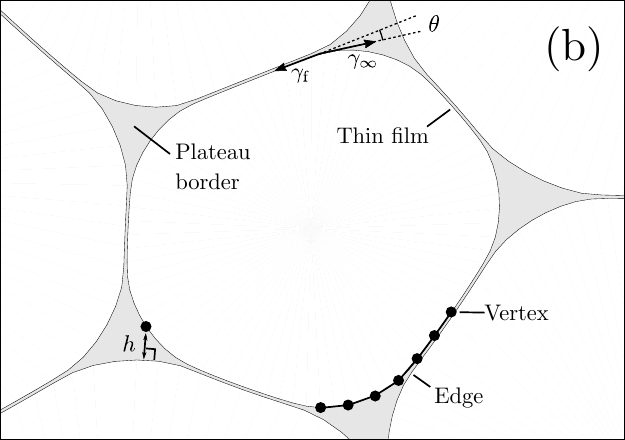}
		\caption{Simulated foam structure for (a)~liquid fraction~$\phi = 10\%$ and no bubble attraction, and (b)~$\phi = 3\%$ and contact angle~$\theta = 10^\circ$, taken from our simulations. The components of the foam are labelled, along with the interface discretization (schematically), local film thickness~$h$ at a vertex, and $\theta$. The surface tension in the films and Plateau borders is~$\gamma_\text{f}$ and~$\gamma_\infty$, respectively, with $\gamma_\text{f} = \gamma_\infty \cos\theta$ \cite{2020-langevin-part}. \label{fig:foam-struct}}
	\end{figure}

	Aqueous foams are packings of gas bubbles in liquid, as illustrated in Fig.~\ref{fig:foam-struct}. They have elasticity from the surface tension of the interfaces, and plasticity due to bubble rearrangements~\cite{1999-weaire}. A foam's properties hence differ substantially from those of its components~\cite{2013-cantat}, and it is a model rheological material whose microstructure is accessible in experiments~\citep{2012-denkov,2023-stewart}. Foams also have a multitude of applications, from foods and drinks~\citep{1999-weaire} to soil treatment~\citep{2016-geraud} and fire suppression~\citep{2012-martin}.
	
	However, foams are unstable, ageing due to coarsening in addition to film breakage \citep{1997-chae} (which we consider no further, noting that it can be suppressed in experiments \cite{2013-roth,2023-pasquet-b}). Coarsening arises from the diffusion of dissolved gas through the liquid, primarily through the thin films between bubbles, and transported from small to large bubbles due to the higher pressures of the former \citep{2017-schimming}. Thus, the mean bubble volume increases as the small bubbles vanish \citep{2010-lambert}. This process may be detrimental in fire-suppression applications, for example, due to the increased mixing of air and fuel vapour within larger bubbles \citep{2012-martin}, and can hasten the perishing of foods \cite{2012-martin-b}. While the dynamics of a confined foam under coarsening differ from bulk foams, the pressure difference required to initiate flow through a porous medium, during soil treatment for example, is also affected by coarsening \citep{2018-jones}. Additionally, foam coarsening is an accessible model for grain growth in crystalline solids \citep{1952-smith}.
	
	Within a coarsening two-dimensional foam with vanishing liquid fraction $\phi$ (the ratio of liquid area to total foam area \cite{2013-cantat}), termed a dry foam, a bubble's growth rate is determined only by its number of neighbours, through von Neumann's law \citep{1952-von-neumann} (discussed in Sec.~\ref{sub:coa:growth-rates}). Experiments and simulations \citep{1992-glazier,1993-stavans} both show that the foam approaches a scaling state, in which statistical distributions scale uniformly with time $t$ \citep{1986-mullins,2010-lambert}. For a given bubble, let $R$ be the radius of a circle with the same area \cite{1988-princen}; i.e.\ its effective radius. It may be shown that the mean effective radius increases as $\langle R \rangle \sim t^{1/2}$ in such a state \citep{1986-mullins}. Similarly, the scaling state of a three-dimensional dry foam is also well established, with the same behaviour of $\langle R \rangle$ \citep{2006-thomas,2010-lambert} (where $R$ is now the radius of a sphere with the same volume), although the bubble growth rates depend upon their precise geometries in this case \citep{2007-macpherson,2013-cantat}.
	
	However, real foams always have nonzero $\phi$, and large values may be encountered in applications such as fire suppression \cite{2011-laundess} and the fabrication of solid foams \cite{2023-galvani,2013-cantat}. For coarsening in the wet limit $\phi \to 1$ (Ostwald ripening \citep{1993-stavans}), a scaling state is obtained with $\langle R \rangle \sim t^{1/3}$ in two and three dimensions \citep[p.\ 77]{2013-cantat}. However, less is known of coarsening at moderate $\phi$. Experiments are difficult to control because drainage of the foam's liquid under gravity occurs on shorter timescales than coarsening \citep{1999-weaire,2021-born}. However, experiments have been performed using diamagnetic levitation of foams \citep{2013-isert}, as well as in microgravity on the International Space Station (ISS) \citep{2021-born,2023-pasquet,2023-pasquet-b,2023-galvani}. A narrow transition was found between the limiting growth exponents $1/2$ and $1/3$ of $\langle R \rangle$, near the jamming transition $\phi_\text{c}$ at which the bubbles (in a foam without bubble attraction) lose contact ($\phi_\text{c} \approx 16\%$ in two dimensions, and $\phi_\text{c} \approx 36\%$ in three) \cite[p.\ 195]{2013-cantat}.
	
	The theory of coarsening at moderate $\phi$ also remains limited. In two dimensions, bubble growth laws have been developed for foams satisfying the decoration theorem \citep{1991-bolton,2013-roth,2017-schimming} ($\phi \lesssim 3\%$ \citep{2021-jing}), and have been compared with experiments in Hele-Shaw cells \citep{2013-roth,2021-chieco} (accounting for the Plateau borders along the bounding plates). An interpolation between the known growth laws for zero and large $\phi$ was proposed, and found, when averaged, to agree with simulations at intermediate $\phi$ \citep{2012-fortuna}. The gas flow rates between adjacent circular or spherical bubbles, present near the jamming transition $\phi_\text{c}$ (in the absence of bubble attraction), have also been derived \citep{2017-schimming}. The latter work has been extended to three-dimensional bubbles with films, although the growth rates of individual bubbles are not predicted \cite{2023-durian-pre}. We are not aware of any general and rigorous growth laws for $0 < \phi < \phi_\text{c}$, in either two or three dimensions (when the decoration theorem does not apply, in the former case).
	
	Several simulations of coarsening in wet foams have been performed \citep{1991-bolton,2000-gardiner,2012-fortuna,2015-thomas,2018-khakalo}. Evidence for scaling states has been found over a range of $\phi$, in two and three dimensions. However, the Potts model simulations \citep{2012-fortuna,2015-thomas} predict a different form for the transition of the growth exponent of $\langle R \rangle$ with $\phi$, compared with experiments \citep{2013-isert,2023-pasquet-b} and simulations \citep{2018-khakalo} using the bubble model \citep{1995-durian}. While the latter simulations qualitatively reproduce the experimental transition (albeit in two dimensions), the relative rate of diffusion across films and Plateau borders is not predicted.
	
	Most numerical studies have used models which are suited to simulating large numbers of bubbles ($3000$ to $10000$ for the bubble model \citep{2018-khakalo}, and about $10^5$ for the Potts model \citep{2012-fortuna,2015-thomas}), but which do not include accurate bubble deformation. A bubble's growth rate depends on the portion of its surface in contact with other bubbles, along with its pressure and those of its neighbours \cite{2013-roth}, each of which being determined by the bubble geometry (the pressures via the Young-Laplace law, which relates them to interface curvature \citep{1999-weaire}). To our knowledge, only \citeauthor{1991-bolton} \citep{1991-bolton}, \citeauthor{2015-benzi} \citep{2015-benzi}, and \citeauthor{2019-pelusi} \citep{2019-pelusi} have performed numerical coarsening studies that accurately model the bubble shapes in wet foams. The first study is limited to small $\phi$, while the latter two, which used a Lattice-Boltzmann approach, were primarily concerned with bubble rearrangements during coarsening.
	
	Furthermore, the effect of the foam's contact angle $\theta$ on coarsening has not yet been widely investigated. As illustrated in Fig.~\ref{fig:foam-struct}, this is the angle between the tangents of the film and Plateau border interfaces where they meet \citep{1975-toshev-b,1995-denkov}, which arises from an imbalance in their surface tensions \cite[p.\ 49]{2013-cantat}. The contact angle is determined by the surfactant, via the disjoining pressure \citep{2020-langevin-part}, and sets the degree of attraction between bubbles (which increases with $\theta$) \cite{1983-princen}. While $\theta$ may be negligible in typical foams \citep{2021-hohler}, a nonzero contact angle is thought to have affected the results of the ISS coarsening experiments, by delaying the transition in growth exponents to $\phi > \phi_\text{c}$ \citep{2023-pasquet-b}. Additionally, contact angles occur in emulsions \cite{1993-bibette}, to which we expect our results also apply, due to their similar structure to foams \cite{1999-weaire}. Prior work has been done to characterize foam structure at $\theta > 0$ \citep{2018-cox,2021-jing,2021-feng,2023-jing}, along with foam rheology \citep{2016-menon}. We are not aware of numerical studies investigating the effects of $\theta > 0$ on coarsening, although nonzero $\theta$ has been used for technical reasons \cite{2012-fortuna,2015-thomas}.
	
	In this article, we present a quasi-static numerical model of coarsening in two-dimensional wet foams, which accurately models the bubble geometries, and allows $\theta > 0$. Two-dimensional foams are widely studied as more tractable models of the real three-dimensional systems \citep{2014-kahara,2018-cox,2018-khakalo}, and can be approximately realized in Hele-Shaw cells \cite{2013-roth}. Our approach to the foam structure is adapted from the models of \citeauthor{2014-kahara} \citep{2014-kahara} and \citeauthor{2018-boromand} \citep{2018-boromand}, the latter having been widely applied, including to foams and emulsions \citep{2019-boromand,2021-golovkova}. These methods are suited to relatively small systems, with about $1700$ bubbles or fewer. We implement our simulations in Kenneth Brakke's {\sc Surface Evolver} software \citep{1992-brakke,2013-brakke}.
	
	We vary $\theta$ by altering the attractive component of the disjoining pressure. Droplet attraction in emulsions has previously \citep{2021-golovkova} been implemented in the model of \citeauthor{2018-boromand} \citep{2018-boromand} using a similar approach.
	
	Our coarsening model is inspired by that applied analytically by \citeauthor{2008-marchalot} \cite{2008-marchalot} and \citeauthor{2017-schimming} \citep{2017-schimming}, and approximates the gas flow through both the liquid films and the Plateau borders.
	
	Our guiding assumption is that results from accurately-modelled small systems (for which the scaling state is likely inaccessible \citep{2006-thomas}) may give insight into coarsening in macroscopic foams, including by refining the necessarily-coarse approximations used in simulations of large systems.
	
	We describe our numerical methods in Sec.~\ref{sec:methods}. Section~\ref{sec:coarsen-poly} gives our results from simulating two-dimensional disordered foams at various $\phi$ and $\theta$, and we conclude in Sec.~\ref{sec:conclusions}. Instead of simulating the time evolution of foams under coarsening, we consider the instantaneous bubble growth rates, and related properties, in foams at structural equilibrium, due to our relatively small system sizes of $256$ to $1024$ bubbles. Analytical approximations are developed to assist with interpretation, and we give a model for the bubble film lengths in the Appendix.
	
	\vspace{-1em}

	\section{Numerical methods \label{sec:methods}}
	
	We begin by giving our methods for equilibrating foams, before defining our coarsening model, and our approach for generating the initial foam structure.
	
	\vspace{-1em}

	\subsection{Structural model \label{sub:met:struct}}
	
	\subsubsection{Discretization \label{ssb:stru:discretization}}
	
	The foam structure is modelled using the approach of \citeauthor{2014-kahara} \cite{2014-kahara} and \citeauthor{2018-boromand} \cite{2018-boromand}. The bubbles are bounded by a closed interface, with arbitrary shape, of vertices connected by straight edges. This is illustrated schematically in Fig.~\ref{fig:foam-struct}. Hence, all liquid/gas interfaces are explicitly included --- the bubbles are disconnected, and their rearrangements occur without adjustments in the discretization.
	
	The bubble areas are fixed, unlike in the models of \citeauthor{2014-kahara} \cite{2014-kahara} and \citeauthor{2018-boromand} \cite{2018-boromand}, taking the foam's gas to be effectively incompressible \cite[p.\ 27]{2013-cantat}. The liquid is identified with the region outside the bubbles. Since coarsening occurs on substantially longer timescales than structural equilibration \cite{2015-thomas}, we use a quasi-static approach like \citeauthor{2018-boromand} \cite{2018-boromand}. This contrasts with \citeauthor{2014-kahara} \cite{2014-kahara}, who applied their model to flowing foams. We also neglect gravity, and therefore choose to measure all pressures relative to the uniform pressure of the liquid (thus taken to be zero).
	
	Another approach \cite{1992-bolton,2015-jing,2018-cox} is instead to assume that the films separating bubbles have zero thickness, and to treat them separately from the Plateau border interfaces. While this likely has advantages in numerical efficiency, since the interfaces are all circular arcs \cite{1992-bolton}, handling bubble rearrangements in this model is complex for wet foams \cite{2018-cox}, particularly in three dimensions \cite[p.\ 80]{1999-weaire}. Furthermore, a small amount of bubble adhesion is often needed for numerical stability \cite{2015-jing,2018-cox}, and the criteria for adding or removing a film between slightly contacting bubbles could induce artefacts near the jamming transition.
	
	However, a disadvantage of the model we use is that, for numerical convergence, the film thickness must be substantially larger (around $\langle R \rangle / 100$) than in real foams \cite{2013-cantat}. This thickness is set by the interactions between different interfaces, which we now describe.
	
	\subsubsection{Disjoining pressure \label{ssb:stru:dis-press}}
	
	Similarly to \citeauthor{2014-kahara} \cite{2014-kahara}, we implement bubble interactions through a disjoining pressure acting between nearby liquid/gas interfaces. This ensures that bubbles do not overlap. In order to allow $\theta$ to be varied, we select a disjoining pressure inspired by Derjaguin-Landau-Vervey-Overbeck (DLVO) theory \cite[p.\ 94]{2013-cantat}. This includes an exponential electrostatic repulsion term dominant at small interface separations $h$ (see Fig.~\ref{fig:foam-struct}), and a van der Waals attraction which dominates at larger $h$ \cite{2020-langevin-part}. Since we expect transient bubble overlaps ($h \leq 0$) during structural relaxation, we alter the form of the second interaction from $1/h^3$ \cite[p.\ 92]{2013-cantat} to $h e^{-h}$ to improve stability. Hence, we use the disjoining pressure
	\begin{align}
		\Pi_\text{D}(h) = A (1 - \alpha h) \, e^{-h / h_0} , \label{eqn:dispress-form}
	\end{align}
	where $A$, $\alpha$ and $h_0$ are positive constants. Positive disjoining pressure corresponds to interface repulsion. Our aim in selecting this form is a qualitative model for $\Pi_\text{D}$, which allows arbitrary $\theta$ to be set via the constant $\alpha$ (i.e.\ by varying the relative strengths of the attractive and repulsive components). We note that the real functional form of $\Pi_\text{D}$ depends upon the surfactant \cite{2020-langevin-part}.
	
	We are not aware of prior studies using this form for $\Pi_\text{D}$. \citeauthor{2014-kahara} \cite{2014-kahara} and \citeauthor{2018-boromand} \cite{2018-boromand} use a repulsive harmonic interaction. \citeauthor{2021-golovkova} \cite{2021-golovkova} incorporate attractive interactions, but use a piecewise-linear form for $\Pi_\text{D}$ without relating its attractive components to a contact angle.
	
	The constants $A$ and $\alpha$ in Eq.~\eqref{eqn:dispress-form} are set as follows. We enforce that the equilibrium film thickness $h_0$, which is a parameter of the simulations, is attained by a film separating two bubbles whose pressures equal the foam's capillary pressure $\Pi_\text{C}$. This is the area-weighted mean bubble pressure \cite{2021-hohler}. Hence, $\Pi_\text{D}(h_0) = \Pi_\text{C}$ by balancing the pressures on the (flat) film interfaces \cite{1975-toshev}. Films separating bubbles with different pressures will have different equilibrium thicknesses \cite{1988-princen}, as discussed later.
	
	The remaining degree of freedom is used to obtain the desired $\theta$. Let $\gamma(h)$ be the surface tension (in dimensions of force, since our model is two dimensional) of a liquid/gas interface in a flat film of thickness $h$, and let $\gamma_\infty$ be the tension of an isolated interface outside a film. Then \cite{2020-langevin-part}
	\begin{align}
		\gamma(h) &= \gamma_\infty + \frac{1}{2} \int_h^\infty \Pi_\text{D}(s) \, d s , 
			\label{eqn:film-int-tension} \\
		\cos\theta &= \gamma(h) / \gamma_\infty . \label{eqn:contact-angle}
	\end{align}
	An interesting consequence of these equations is that $\Pi_\text{D}$ must have an attractive component even for $\theta = 0$ \cite{1975-toshev-b}, so that $\gamma(h) = \gamma_\infty$ (since $\Pi_\text{D}(h) > 0$ in an equilibrium film \cite[p.\ 96]{2013-cantat}).
	
	The contact angle $\theta_\text{m}$, measured by approximating the simulated bubble interfaces by a collection of circular arcs \cite{1985-kralchevsky-b,1995-denkov}, differs from the value of $\theta$ expected from Eq.~\eqref{eqn:contact-angle} with $h = h_0$. When $\theta = 0$, we find $\theta_\text{m} \approx 3^\circ$ for $\phi \approx \phi_\text{c}$, which manifests the attractive component of $\Pi_\text{D}$ in this case, while the discrepancy decreases as $\theta$ increases. As discussed further in Sec.~\ref{sub:coa:growth-rates}, we find that $h > h_0$ near the jamming transition. Hence, from Eq.~\eqref{eqn:film-int-tension}, and the short-range repulsion in $\Pi_\text{D}$, Eq.~\eqref{eqn:contact-angle} then predicts a larger contact angle. We parametrize our foams with respect to $\theta$ for $h = h_0$, which we recall determines the interface tension in a flat film between bubbles with the capillary pressure.
	
\begin{figure}
		\includegraphics[width=1.0\columnwidth]{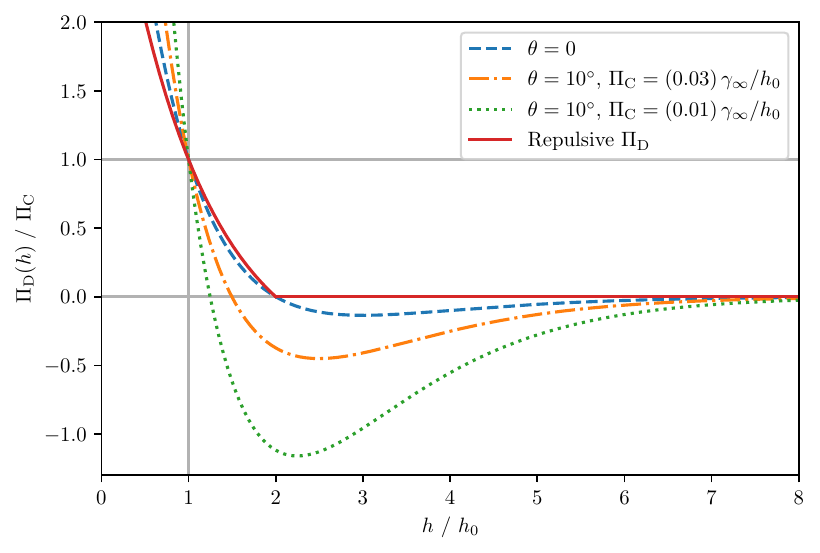}
		\caption{The form of disjoining pressure $\Pi_\text{D}$ we use [Eqs.~\eqref{eqn:dis-press} and~\eqref{eqn:dis-press-rep}], versus film thickness $h$ (relative to its equilibrium value $h_0$). For $\theta > 0$, the capillary pressure $\Pi_\text{C}$ does not scale out. The larger value is representative of a simulated foam with $\phi = 2\%$, and the smaller for a flocculated foam with larger $\phi$ (see Sec.~\ref{sec:coarsen-poly}). \label{fig:dis-press}}
	\end{figure}

	By substituting Eq.~\eqref{eqn:dispress-form} into Eqs.~\eqref{eqn:film-int-tension} and~\eqref{eqn:contact-angle}, and applying the above condition $\Pi_\text{D}(h_0) = \Pi_\text{C}$, our disjoining pressure is
	\begin{align}
		\Pi_\text{D}(h) &= \frac{2 \gamma_\infty}{h_0} \left[\left(1 - \cos\theta
			+ \frac{\Pi_\text{C} h_0}{\gamma_\infty}\right)\right. \nonumber \\
		&\qquad \left. - \left(1 - \cos\theta
			+ \frac{\Pi_\text{C} h_0}{2 \gamma_\infty}\right) \frac{h}{h_0}\right]
			e^{1 - h / h_0} . \label{eqn:dis-press}
	\end{align}
	In order to model foams with no bubble attraction, we also implement a repulsive $\Pi_\text{D}$. This is obtained by setting $\alpha = 0$ in Eq.~\eqref{eqn:dispress-form}, along with a cutoff such that $\Pi_\text{D}(h) = 0$ for $h > 2 h_0$ (so bubble neighbours can be reliably calculated), and adjusted so $\Pi_\text{D}(h)$ is continuous while still satisfying $\Pi_\text{D}(h_0) = \Pi_\text{C}$:
	\begin{align}
		\Pi_\text{D}(h) &= \dfrac{\Pi_\text{C}}{e - 1} \begin{cases}
				e^{2 - h / h_0} - 1 , & \text{for $h \leq 2 h_0$} ; \\
				0, & \text{otherwise} .
			\end{cases} \label{eqn:dis-press-rep}
	\end{align}
	We note that $\gamma(h) > \gamma_\infty$ and $\theta$ is undefined (inevitable for a repulsive $\Pi_\text{D}$), by Eqs.~\eqref{eqn:film-int-tension} and~\eqref{eqn:contact-angle}. Equations~\eqref{eqn:dis-press} and~\eqref{eqn:dis-press-rep} are plotted in Fig.~\ref{fig:dis-press}.
	
	Equation~\eqref{eqn:film-int-tension} is not exact for curved interfaces, which are ubiquitous in foams, and a realistic disjoining pressure would depend upon the interface curvature \cite{1995-denkov}. However, we neglect these effects as a simplifying assumption. This is the Derjaguin approximation, justified when the radii of curvature of the interfaces are large compared to their separation $h$ \cite{1995-denkov}. The latter assumption does not hold in general for our simulations, although we expect that it does for the film interfaces.
	
	We now explain further the requirement for larger $h_0$ in our simulations than in real foams. It appears necessary for convergence that $\Pi_\text{D}$ not vary too rapidly with $h$ near $h_0$. In principle, $h_0$ could be set arbitrarily small, but, to avoid unphysical interface overlaps at equilibrium, $|\Pi_\text{D}'(h_0)|$ would need to be correspondingly large (recalling that variations in bubble pressure correspond to variations in the value of $\Pi_\text{D}$ in equilibrium films). Furthermore, even given our larger $h_0$, we believe $|\Pi_\text{D}'(h_0)|$ is much smaller than found in real foams \cite{1992-bergeron}, giving rise to larger variations in film thickness between bubbles of different pressure (although some variations are expected in real foams \cite{1988-princen}). We discuss this further in Sec.~\ref{sub:coa:growth-rates}.
	
	The disjoining pressure $\Pi_\text{D}(h)$ is applied to each vertex on the liquid/gas interfaces (unlike \citeauthor{2014-kahara} \cite{2014-kahara} and \citeauthor{2018-boromand} \cite{2018-boromand} who apply it to the edges in their foam models), with $h$ equal to the shortest distance to another interface (the local film thickness), as shown in Fig.~\ref{fig:foam-struct}.
	
	\subsubsection{Vertex neighbours \label{ssb:stru:neigh}}
	
\begin{figure}
		\includegraphics[width=1.0\columnwidth]{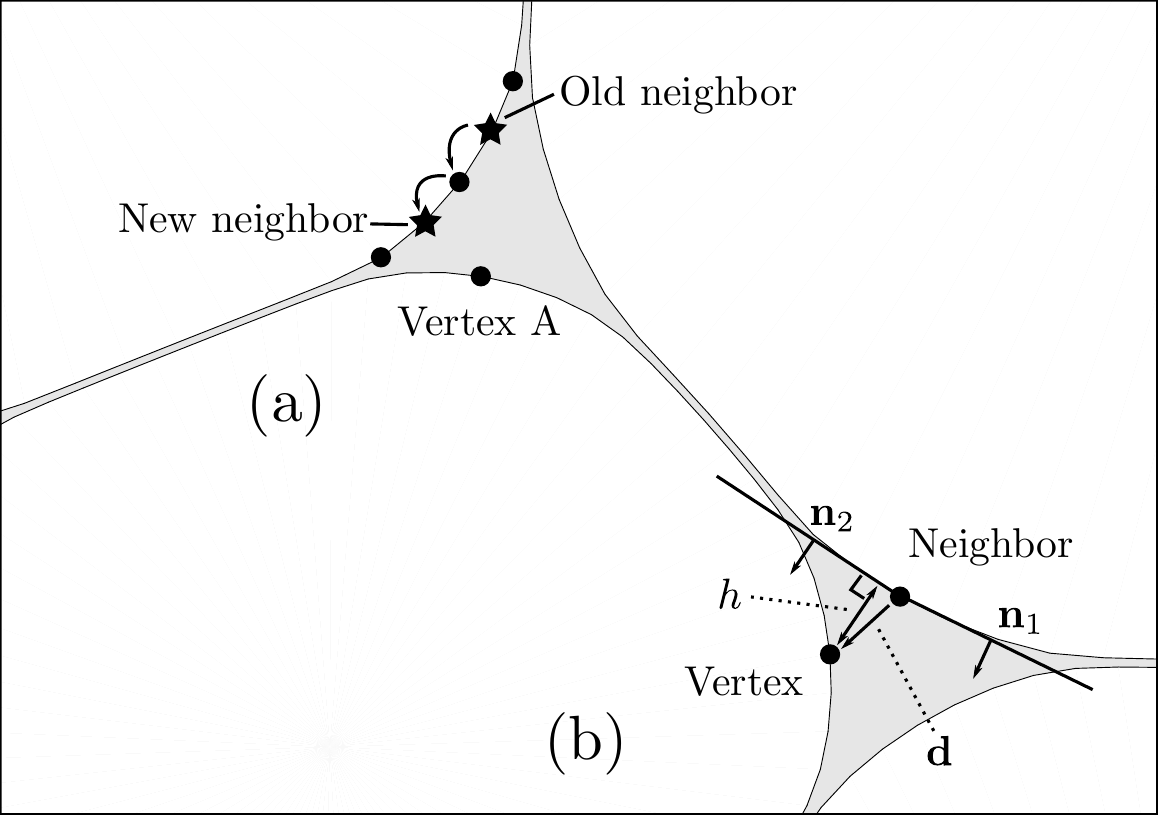}
		\caption{Schematic of (a) the local neighbour search, and (b) the piecewise-linear interface extrapolation. The outward unit normals to the edges adjoining the neighbouring vertex in (b) are $\mathbf{n}_1$ and $\mathbf{n}_2$, the displacement of the vertex from its neighbour is $\mathbf{d}$, and $h$ is the extrapolated shortest distance from the vertex to an opposing interface. \label{fig:neighbour-search-and-extrap}}
	\end{figure}

	To determine the local film thickness $h$ at a vertex~$\text{A}$, neighbour searches are first performed to find the vertex closest to $\text{A}$ which lies on a different interface, illustrated in Fig.~\ref{fig:neighbour-search-and-extrap}(a).
	
	In order to improve the efficiency of these searches, compared to a brute-force approach, we use the fact that all vertices lie on particular bubble interfaces, and their adjoining vertices thereon do not change. Our nearest-neighbour algorithm is:
	
	(1)~Cover each bubble with a circle whose centre is the centroid of its vertices, and whose radius is the minimum required to cover each vertex.
	
\begin{figure}
		\includegraphics[width=1.0\columnwidth]{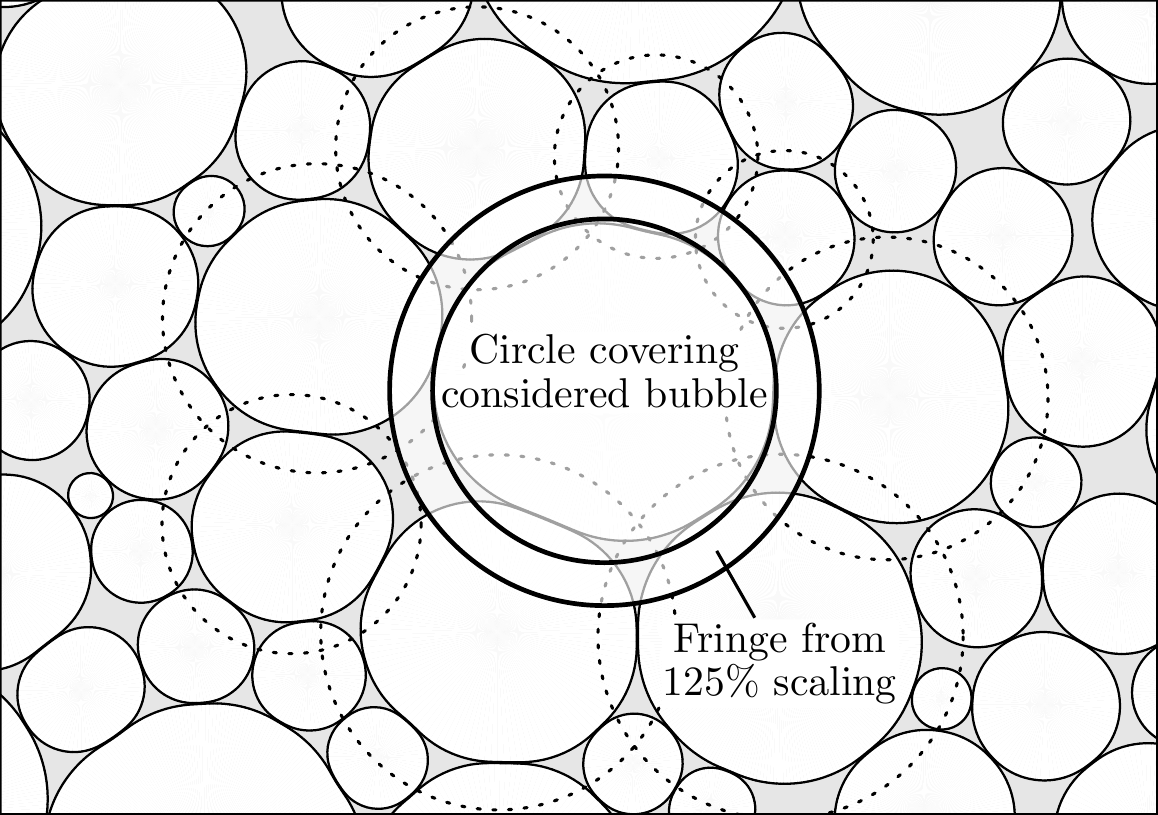}
		\caption{Schematic of the covering circles used to determine neighbouring bubbles for the purpose of finding vertex neighbours. The circle for a particular bubble is shown, including its fringe used to improve numerical stability, along with the circles of its neighbouring bubbles (i.e.\ those that overlap the first circle). \label{fig:bubble-neighbours}}
	\end{figure}

	(2)~For each bubble, determine the bubbles whose circles overlap its own, illustrated in Fig.~\ref{fig:bubble-neighbours}. These are the neighbouring bubbles, and are found by calculating the distance between each circle centre, since our systems contain relatively few bubbles (no more than $1024$). For these comparisons, the circle radii are scaled by $125\%$ (an arbitrary value, but sufficiently large). If a scaling is not performed, then interface overlaps can occur during bubble rearrangements, as some bubbles would intersect one another before registering as neighbours.
	
	(3)~For each vertex~$\text{A}$ in each bubble, calculate the closest vertex in each neighbouring bubble. If the latter bubble was a neighbour during the last pass of the algorithm, then the previously-closest vertex provides an initial guess. Otherwise, we take the vertex closest to the centroid of the bubble on which vertex~$\text{A}$ lies (since this need only be found once for all vertices on this bubble). The closest vertex is then obtained through a local search along the neighbouring bubble's interface, starting at the guessed vertex, and, at each step, moving to an adjoining vertex if it is closer to~$\text{A}$. This is illustrated in Fig.~\ref{fig:neighbour-search-and-extrap}(a). The process is only guaranteed to give a vertex at a local minimum of distance, but works well in practice due to the smooth bubble geometries (see Fig.~\ref{fig:foam-struct}) and the choice of initial guess.
	
	(4)~For each vertex~$\text{A}$ in each bubble, compare the distance to the closest vertex in each neighbouring bubble. The closest of these is then vertex~$\text{A}$'s nearest neighbour.
	
	We are not aware of descriptions of the nearest-neighbour approaches used in prior simulations of this type, although we note that \citeauthor{2023-okuda} \cite{2023-okuda} use octrees in comparable three-dimensional simulations of biological cells~\cite{2023-okuda-b}. The efficiency of these algorithms appears to be of importance in scaling such foam simulations to larger systems and higher mesh refinement.
	
	\subsubsection{Interface extrapolation \label{ssb:stru:extrap}}
	
	Having determined the nearest neighbour of each vertex, we then calculate the local film thickness $h$. We could set $h$ equal to the distance from the vertex to its neighbour, but this would result in an unrealistic roughness in the interfaces --- this approach has previously been used to model static friction in other materials \cite{2018-boromand}.
	
	Instead, we use a piecewise-linear interface extrapolation, such that $h$ is the minimum distance to the half-infinite extensions of the two edges adjoining the neighbour vertex, as illustrated in Fig.~\ref{fig:neighbour-search-and-extrap}(b). Let $\mathbf{n}_1$ and $\mathbf{n}_2$ be outward unit normals to the edges, let $\mathbf{d}$ be the displacement of the considered vertex from its neighbour, and let $a = 1$ if the interface is convex at the neighbour vertex [as in Fig.~\ref{fig:neighbour-search-and-extrap}(b)] and $a = -1$ if it is concave. Then
	\begin{align}
		h = a \, \min \Big[\max\left(a \, \mathbf{d} \cdot \mathbf{n}_1,
			a \, \mathbf{d} \cdot \mathbf{n}_2 \right), |\mathbf{d}| \Big] .
			\label{eqn:interface-extrap}
	\end{align}
	This ensures that $h < 0$ if the vertex overlaps the neighbour interface during structural relaxation, so the disjoining pressure acts to oppose the overlap.
	
	Issues of interface extrapolation or interpolation also arise in other fluid dynamics simulations. For example, \citeauthor{2004-bazhlekov} \cite{2004-bazhlekov} use a spherical interpolation between vertices to obtain the local film thickness in three-dimensional boundary integral simulations.
	
	We use the above extrapolation due to its closeness to the geometry of the discretization. Other approaches might allow improved convergence at lower mesh refinement. \citeauthor{2014-kahara} \cite{2014-kahara} and \citeauthor{2018-boromand} \cite{2018-boromand} bypass the issue of interface extrapolation by explicitly calculating the shortest distance between edges. However, we determine neighbouring vertices, rather than neighbouring edges, under the assumption of greater efficiency, due to the simpler distance calculations involved.
	
	\subsubsection{Implementation \label{ssb:stru:implement}}
	
	We implement our simulations using the {\sc Surface Evolver}, developed by Kenneth Brakke \cite{1992-brakke,2013-brakke}. This software is frequently applied in the study of foams \cite{2003-kraynik,2015-jing,2021-hohler}, usually under the assumption that the liquid films have zero thickness.
	
	The {\sc Surface Evolver} may be extended using its scripting language. We have implemented finite film thickness in this manner, via a disjoining pressure and neighbour searches. Our scripts are not compiled, so it is likely that our simulations could be made faster by implementing the routines in the software's public source code. We also do not take advantage of any significant parallel processing.
	
	We note that pairwise repulsion between vertices already exists in the {\sc Surface Evolver} as `knot' energies~\cite{2013-brakke}. However, it is not straightforward to adapt these to our purposes, due to our desire for interface extrapolation and interactions only between nearest neighbours.
	
	As is usual in the {\sc Surface Evolver}, local energy minimization is used to obtain the equilibrium foam structures. Let $\Gamma$ be the union of all liquid/gas interfaces. Then the foam's total energy (recalling that its liquid and gas are treated as incompressible) is given by
	\begin{align}
		E = \int_\Gamma \gamma(h) \, d l , \label{eqn:total-energy}
	\end{align}
	where $\gamma(h)$ is obtained from Eq.~\eqref{eqn:film-int-tension}, and $h$ is the local film thickness from Eq.~\eqref{eqn:interface-extrap} (which varies around $\Gamma$). This quantity is minimized using standard conjugate gradient iterations, as implemented in the {\sc Surface Evolver} \cite{1992-brakke,2013-brakke}. Let $l_i$ be the length of interface associated with the $i^\text{th}$ vertex (half the length of its adjoining edges), and let $h_i$ be the corresponding local film thickness. Also, let $L_j$ be the length of the $j^\text{th}$ edge. Then $E$ is expressed in the simulations as
	\begin{align}
		E = \gamma_\infty \sum_{\text{edge $j$}} L_j + \frac{1}{2} \sum_{\text{vertex $i$}} l_i \int_{h_i}^\infty \Pi_\text{D}(s)
			\, d s ; \label{eqn:energy-implemented}
	\end{align}
	where the integrals are determined explicitly from Eq.~\eqref{eqn:dis-press} or~\eqref{eqn:dis-press-rep}. As an approximation, we neglect the contribution to $\bm{\nabla} E$ from variations in $l_i$, corresponding to corrections to the surface tension forces from $\Pi_\text{D}$ (expected to increase with $\theta$). Since our implementation doesn't allow derivatives in neighbour coordinates, we also double the second term in Eq.~\eqref{eqn:energy-implemented} when calculating $\bm{\nabla} E$. This is equivalent to including such derivatives, under the approximation that each vertex is the nearest neighbour of its own nearest neighbour, and that the corresponding distances are those between the vertices.
	
	Because the bubbles are expected to rearrange slowly, we perform the time-consuming neighbour searches every $20$ conjugate gradient iterations. However, the thicknesses $h_i$ are recalculated before each iteration, using the current coordinates of the vertex neighbours.
	
	The edges on each bubble interface are kept at an approximately uniform length by using the {\sc Surface Evolver}'s vertex averaging routines, in addition to restricting vertices to move only in the direction normal to their interface. \citeauthor{2014-kahara} \cite{2014-kahara} instead implemented elasticity in their edges.
	
	Transiently, and usually during the early stages of structural relaxation, the assumptions of the interface extrapolation may fail. For example, spikes in the discretization of a bubble's interface may result in a large interface overlap ($h \ll 0$) being falsely registered for a vertex. This would cause the vertex to experience an extremely large repulsive force. To avoid this, we detect instances of large overlap, and set the coordinates of the overlapping vertex and its neighbour to the mean of those adjoining them on their interfaces. We also detect spikes using the angle turned by successive edges on an interface, suppressing them by the same method.
	
	Our use of the {\sc Surface Evolver}'s relaxation routines to implement vertex interactions have resulted in the following caveat: the vertex neighbour properties are not recalculated during a given conjugate gradient iteration, interfering with the selection of a step size \cite{1992-brakke}. This slows relaxation, and means the energy is no longer guaranteed to decrease. In practice, we find that this method is still considerably faster than gradient descent with a fixed time step (to which the caveat does not apply), and that systematic energy increases are rare for our system parameters (with their occurrence also being reduced at higher mesh refinement). We tried implementing our vertex interactions in the {\sc Surface Evolver} source code to allow an exact use of the conjugate gradient algorithm, although we found the result unusable due to the step size decreasing to zero prior to convergence being clearly established. This may be due to an incompatibility between the system energy and its calculated gradient, noting that our vertex interactions are asymmetric, but remains unresolved. Therefore, we use the first approach in the results reported here.
	
	We have compared foams equilibrated by each of the above techniques (for $256$ bubbles, $\phi = 2\%$, and $\theta = 10^\circ$), and find them consistent, with the first method resulting in the lowest final energy (this energy differs by less than $0.002 \%$ among the techniques). Since our source-code implementation includes contributions to $\bm{\nabla} E$ from variations in the $l_i$ of Eq.~\eqref{eqn:energy-implemented} (recalling that we expect these contributions to increase with $\theta$), this also tests our neglect of these in the technique we employ here (although we note the bubble pressures are slightly increased with this latter method, by about $3\%$).

	\subsection{Coarsening model \label{sub:met:coarsening}}
	
	Having given our methods for structural relaxation, we now describe our coarsening model.
	
	Coarsening occurs due to the transport of dissolved gas through the foam's liquid \cite{2013-cantat}. Let $c(\bm{r}, t)$ be the gas concentration at position $\bm{r}$ in the liquid, and at time $t$. Following \citeauthor{2017-schimming} \cite{2017-schimming}, we assume that almost all the gas transfer between bubbles occurs when the gas concentration is close to equilibrium. Hence, $c$ satisfies Laplace's equation
	\begin{align}
		\nabla^2 c = 0 . \label{eqn:laplace-eqn}
	\end{align}
	Boundary conditions are given by Henry's law $c = H p$ on interface of each bubble, where $p$ is the bubble's pressure and $H$ a constant related to the solubility \cite[p.\ 109]{2013-cantat}. Only differences in $c$ are relevant, so pressures relative to the liquid may be used.
	
\begin{figure}
		\includegraphics[width=1.0\columnwidth]{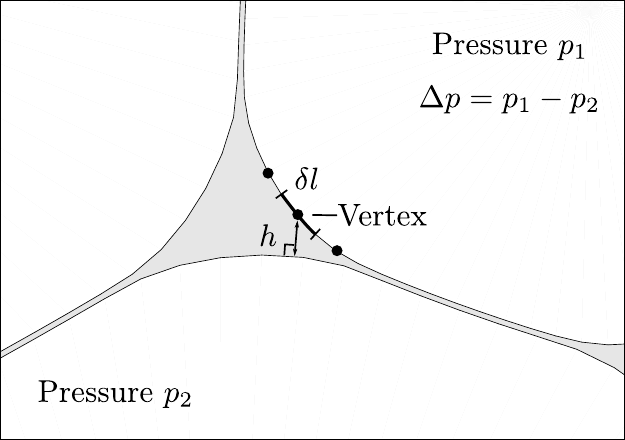}
		\caption{Illustration of the quantities used in Eq.~\eqref{eqn:approx-flow-rate}. With these definitions, $\delta F$ is a contribution to the loss of gas from the bubble with pressure $p_1$. \label{fig:coarsening}}
	\end{figure}

	Let $\delta F$ be the gas flow rate (in dimensions of area per time) across an element $\delta l$ of a bubble's interface, due to a nearby bubble. Using the approach of \citeauthor{2008-marchalot} \cite{2008-marchalot} and \citeauthor{2017-schimming} \cite{2017-schimming}, we approximate this rate as that between two infinite, parallel, straight interfaces, separated by the local film thickness $h$ (the same as used to determine $\Pi_\text{D}$). Let the pressure difference between the two bubbles be $\Delta p$. The solution of Laplace's equation is linear between such interfaces, so \cite[p.\ 109]{2013-cantat}
	\begin{align}
		\delta F \approx D \, \delta l \, \Delta p / h ,
			\label{eqn:approx-flow-rate}
	\end{align}
	where $D$ is a diffusion coefficient (incorporating $H$). The gas flows towards the bubble with lower pressure. We apply this approximation to each vertex, taking $\delta l$ to be half the sum of the lengths of the two adjoining edges. The relevant quantities are illustrated in Fig.~\ref{fig:coarsening}.
	
	By summing $\delta F$ for each vertex on a bubble's interface, we obtain an approximation to its area growth rate $\dot{A}$, which accounts for gas transfer through both the thin films and Plateau borders.
	
	Both contributions were previously included in Lattice-Boltzmann simulations~\cite{2015-benzi,2019-pelusi}, although these studies were focused on the motion of bubbles during coarsening, rather than their growth rates. The contributions were also present in the phase-field simulations of \citeauthor{2002-fan} \cite{2002-fan}. However, their focus was not on foam-like systems. Otherwise, to our knowledge, prior simulations of wet foams with accurate bubble geometries have only included gas flow through the films \cite{1991-bolton} (this is the border blocking assumption \cite{2013-roth}). Simulations using simplified bubble geometries have implemented approximate contributions of both types \cite{2012-fortuna,2015-thomas,2018-khakalo}, although more work is needed to justify their forms for the Plateau border contributions, which were originally derived for very large $\phi$.
	
	\citeauthor{2017-schimming} \cite{2017-schimming} compared their approximate gas flow rates through Plateau borders, obtained by a similar approach to the above, with those given by solving Laplace's equation numerically. They obtained close agreement for thin films. However, our simulations use relatively thick films to ensure convergence, and $\delta l$ is finite (around one hundredth of the bubble perimeter), whereas $\delta l \to 0$ in their analytical results. Therefore, the error in our bubble growth rates is not fully characterized. We also note that each element of interface can exchange gas with only one neighbouring bubble, which may induce further errors in the gas flow through Plateau borders, where neighbouring bubbles meet~\cite{2022-hohler}. We compare the predictions of our coarsening model with existing growth laws in Sec.~\ref{sec:coarsen-poly}.
	
	\subsection{Generation of disordered foams \label{sub:met:generate}}

	We now describe our process for generating the initial disordered foam structure.
	
	As is usual \cite{2018-cox,2021-jing}, we begin with a Voronoi tessellation of the plane, with a domain containing the desired number of bubbles subject to periodic boundary conditions. These tessellations are obtained using the {\sc vor2fe} software by Kenneth Brakke \cite{1986-brakke}. The resulting dry foam is then equilibrated using standard techniques \cite{1992-brakke}, with one straight edge per liquid film for efficiency.
	
	To obtain a more representative bubble area distribution for coarsening foams than that of a Voronoi tiling, we sample from the area distribution fitted by \citeauthor{2013-roth} \cite{2013-roth} to experimental data for quasi-two-dimensional foams (i.e.\ foams in Hele-Shaw cells). The sampled areas are randomly assigned to the bubbles in the Voronoi foam \cite{2015-jing}. We note that the data of \citeauthor{2013-roth} \cite{2013-roth} are for foams with small $\phi$, and may differ from the truly two-dimensional case we model. Furthermore, neighbour correlations are not introduced.
	
	The sampled areas are more polydisperse than those obtained from a Voronoi tessellation. Let $R_{2 1} \equiv \langle R^2 \rangle / \langle R \rangle$ \cite[p.\ 251]{2013-cantat}. Then the bubble distributions have a representative polydispersity $\sigma = R_{2 1} / \sqrt{\langle R^2 \rangle} - 1$ \cite{2018-cox} of $0.088$ and $0.046$, respectively. However, we have not observed qualitative differences in the foam or bubble properties considered in Sec.~\ref{sec:coarsen-poly} when the latter areas are used. The same relationships between growth rate and radius, for example, are observed, but with less revealed at the lower polydispersity due to the narrower interval of realized radii. Hence, even if the sampled area distribution is unrealistic for larger $\phi$, its high polydispersity remains useful for characterizing the relationships.
	
	Next, the system is annealed following the approach of \citeauthor{2003-kraynik} \cite{2003-kraynik}, adapted to two dimensions, so that the distribution of bubble topology better approximates that in a real foam \cite{2003-kraynik}. By deforming the simulation domain discontinuously, a sequence of extensional strains is applied, with relaxation after every change. The domain is then deformed (using cycles of simple and extensional shear \cite{1999-weaire}) to relax the deviatoric stress to zero (within a tolerance $10^{-5}\,\gamma_\infty / \langle R \rangle$). Stress relaxation of the initial structure was performed in the dry foam coarsening simulations of \citeauthor{1992-herdtle} \cite{1992-herdtle}. We have not investigated the effects of varying this preparation process.
	
	The liquid fraction $\phi$ is then set, by duplicating the edges and vertices associated with each bubble (since these are shared by neighbours in the dry model), and uniformly expanding the simulation domain to $1/(1 - \phi)$ of its original area. The same transformation is applied to the bubble centroids, while the bubble shapes and sizes are maintained. Hence, $\phi$ is set by effectively adding liquid to the system, keeping the gas area fixed \cite{2018-cox}.
	
	We next subdivide all edges on the bubble interfaces $m$ times in order to obtain the desired mesh refinement. Some of the initial edges are very short, so these are skipped to improve convergence. The longer edges are further subdivided to compensate. The average number of edges (and vertices) per bubble is then $6 \times 2^m$, since $6$ is the mean number of bubble neighbours (and hence edges) in the initial dry foam \cite[p.\ 29]{1999-weaire}. After initial structural relaxation, vertices are added or deleted to ensure small and large bubbles have acceptable numbers of vertices.
	
	The foam's structure is relaxed using the approach of Sec.~\ref{sub:met:struct}. We recall, from Eq.~\eqref{eqn:dis-press}, that the strength of $\Pi_\text{D}$ (determining the equilibrium film thickness $h_0$) is set via the foam's capillary pressure $\Pi_\text{C}$, which is not known beforehand. Hence, we initially relax the foam using an estimate of $\Pi_\text{C}$ adapted from \citeauthor{1979-princen} \cite{1979-princen} (i.e.\ given by the result for hexagonal foams, except with $\phi_\text{c}$ shifted to $16\%$, suitable for disordered foams \cite[p.\ 195]{2013-cantat}). A measurement of $\Pi_\text{C}$ in the relaxed foam is used to reset the strength of $\Pi_\text{D}$, and the foam is relaxed again. We find that around $5$ iterations of this process are needed for $\Pi_\text{C}$ to converge to precision $10^{-2} \, \gamma_\infty / L$ (where $L^2$ is the area of the foam's periodic domain). Once this is done, the desired equilibrium film thickness $h_0$ is approximately obtained. However, as mentioned above, the film thickness will vary between bubbles, due to their variations in pressure~\cite{1988-princen}.
	
	Finally, the deviatoric stress is relaxed again, giving an initial wet foam structure that is intended to be representative of a bulk foam. Relaxing the dry foam before setting $\phi$ \cite{2015-jing} is useful for computational efficiency.

	\section{Coarsening in wet foams \label{sec:coarsen-poly}}
	
	We now present our main results, on coarsening in two-dimensional disordered foams for various liquid fractions~$\phi$ and contact angles~$\theta$.
	
	Our simulations follow a similar approach to \citeauthor{1990-bolton} \cite{1990-bolton} and \citeauthor{2018-cox} \cite{2018-cox} (since applied by others \cite{2021-jing,2021-feng,2023-jing}) in gradually increasing $\phi$, and relaxing the foam's structure, film thickness, and applied stress at each step. Hence, we obtain data on the foam structure over a range of $\phi$. We also capture the flocculation \cite{2018-cox} discussed below, which arises differently if $\phi$ is set to a large value in a single step.
	
	We perform these liquid fraction sweeps from $\phi = 2\%$ to $25\%$. The upper bound is arbitrary, and chosen so that plenty of data is obtained for foams beyond the jamming transition ($\phi_\text{c} \approx 16\%$~\cite[p.\ 195]{2013-cantat}). The lower bound is determined by convergence. Previous simulations (with $\theta > 0$) using a different structural model have used the same upper bound \cite{2018-cox,2021-jing}.
	
	Simulations are run for $\theta \in \{0, 2.5^\circ, 5^\circ, 7.5^\circ, 10^\circ\}$, and for repulsive $\Pi_\text{D}$. Our range for $\theta$ is comparable to prior studies~\cite{2021-jing,2021-feng}. Larger $\theta$ results in slower convergence, and requires higher mesh refinement.
	
	We set the equilibrium film thickness $h_0$ to $\langle R \rangle / 100$. Our simulations cannot converge for values much below $\langle R \rangle / 200$, recalling our comments in Sec.~\ref{sub:met:struct}.
	
	The simulated foams contain $256$ or $1024$ bubbles. Our energy-change threshold for halting structural relaxation is $10^{-6} \, \gamma_\infty L$ (over $20$ conjugate gradient iterations), and we use $m = 4$ refinements of the initial dry-foam mesh. Hence, the mean number of vertices per bubble is initially $96$, before the discretization of large or small bubbles is adjusted. The liquid fraction is incremented in steps of $\Delta \phi = 0.5\%$. These parameters have been varied to check convergence. The growth rates at $\phi = 2\%$ for $m = 4$ differ from their values at $m = 5$ by a scaling of about $110\%$, which is uniform for all bubbles to a good approximation.
	
	The execution time for one liquid fraction sweep with $256$ bubbles is about $\unit[100]{hours}$ on a PC with a recent $16$-core Intel i7 processor, $\unit[16]{GB}$ of RAM, and an SSD, recalling that little parallel processing is used. A sweep with $1024$ bubbles takes around $\unit[700]{hours}$. The execution times are usually lower for smaller $\theta$ and larger $\phi$, and when bubble areas are taken directly from the initial Voronoi tessellation.
	
	In the following subsections, we first describe the global properties of the simulated foams. We then discuss coarsening-related properties of individual bubbles, and develop some analytical approximations. Next, we describe the variation of the bubble growth rates with $\phi$. Finally, we consider the aggregated properties of the foam's bubbles, in order to better quantify their variation with $\phi$ and $\theta$.
	
	\subsection{Foam properties \label{sub:coa:equilibrium}}
	
	If $\bar{\bm{\tau}}$ is the stress tensor spatially averaged over the foam, after \citeauthor{1970-batchelor} \cite{1970-batchelor} (and determined by the bubble pressures and interface geometry \cite[p.\ 175]{2013-cantat}), then the osmotic pressure \cite{1979-princen} is given by \cite{1995-hutzler-b,2021-hohler}
	\begin{align}
		\Pi_\text{O} = -\tfrac{1}{2} \Tr \bar{\bm{\tau}} \label{eqn:osmotic-press}
	\end{align}
	in two dimensions, recalling that we set the liquid pressure to zero. $\Pi_\text{O}$ is the average pressure in the foam, above that of the liquid \cite{2021-hohler}. It is also approximately the pressure that a piston would need to exert on the foam to maintain its present $\phi$, where this piston is permeable to the liquid but not the bubbles \cite{1979-princen,2021-hohler}.
	
\begin{figure}
		\includegraphics[width=1.0\columnwidth]{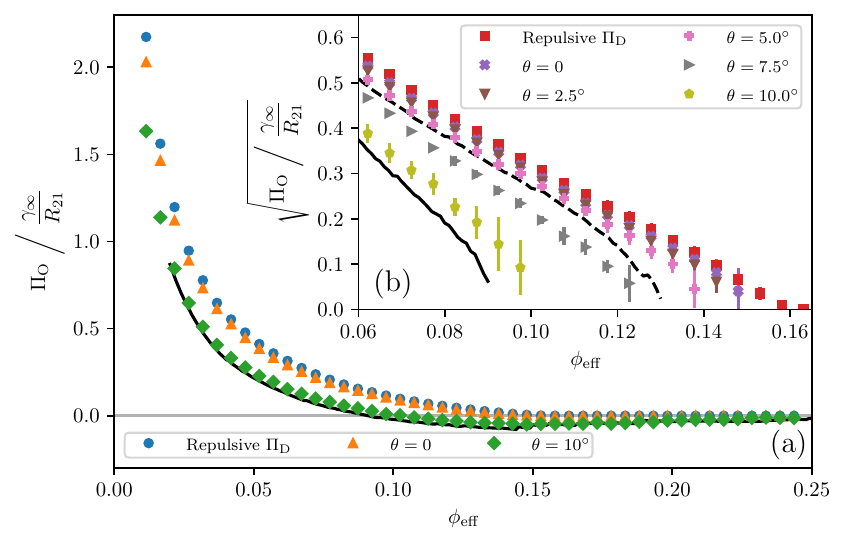}
		\caption{Osmotic pressure~$\Pi_\text{O}$ versus effective liquid fraction (defined in the text) in foams with different contact angles $\theta$. Data is given for $1024$-bubble runs in~(a), while~(b) gives the variation of~$\sqrt{\Pi_\text{O}}$ for $256$-bubble runs to clarify the zero of~$\Pi_\text{O}$. The data in~(b) is averaged over $5$ different initial foams, with the square-root then being taken, and the error bars give the propagated sample standard deviation of the linear data. For comparison, the solid (dashed) curve is for a $1500$-bubble foam with $\theta = 10.8^\circ$ ($\theta = 5.1^\circ$) and $h_0 = 0$, simulated with the model of \citeauthor{2018-cox} \cite{2018-cox}. The curves in (b) end when $\Pi_\text{O} < 0$. \label{fig:osmotic-press-v-liquid-fraction}}
	\end{figure}

	In Fig.~\ref{fig:osmotic-press-v-liquid-fraction}, we plot the simulated variation of $\Pi_\text{O}$ with effective liquid fraction (defined below) for several $\theta$. $\Pi_\text{O}$ is positive for all considered $\theta$ at $\phi$ up to a threshold. This threshold is approximately the jamming transition $\phi_\text{c} \approx 16\%$ \cite[p.\ 195]{2013-cantat} for small $\theta$, as expected \cite{1995-hutzler-b}, and decreases for larger $\theta$, as for hexagonal foams \cite{1979-princen}. For $\phi$ above the threshold, $\Pi_\text{O} \approx 0$, so the foam is no longer under compression. For repulsive $\Pi_\text{D}$ and $\theta = 0$, the variation of $\Pi_\text{O}$ with $\phi$ is in qualitative agreement with prior simulation results \cite{1995-hutzler-b}. From Fig.~\ref{fig:osmotic-press-v-liquid-fraction}(b), we are also consistent with their assertion of a quadratic zero at $\phi = \phi_\text{c}$, for small $\theta$.
	
	Alongside our data, we plot that obtained using the same simulation model as \citeauthor{2018-cox} \cite{2018-cox} (the {\sc Surface Evolver} with $h_0 = 0$). Good agreement is observed. To make the comparison, we correct for our finite film thickness by plotting against the effective liquid fraction $\phi_\text{eff}$, instead of $\phi$, obtained by subtracting from the foam's liquid area a strip of width $h_0/2$ around the perimeter of each bubble \cite{1979-princen,1989-khan}. If $\langle P \rangle$ is the mean bubble perimeter, and $\langle A \rangle$ the mean bubble area, then $\phi_\text{eff} = \phi - (1 - \phi) \, \langle P \rangle h_0 / (2 \langle A \rangle)$. A three-dimensional version was used by \citeauthor{1997-mason} \cite{1997-mason}.
	
\begin{figure}
		\includegraphics[width=0.7\columnwidth]{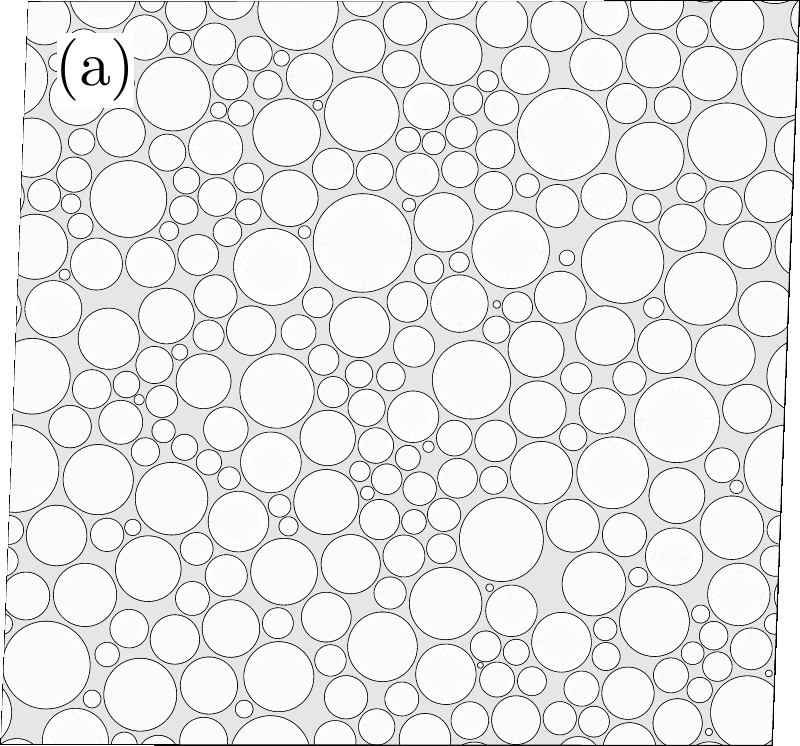}
		
		\vspace{1em}
		\includegraphics[width=0.7\columnwidth]{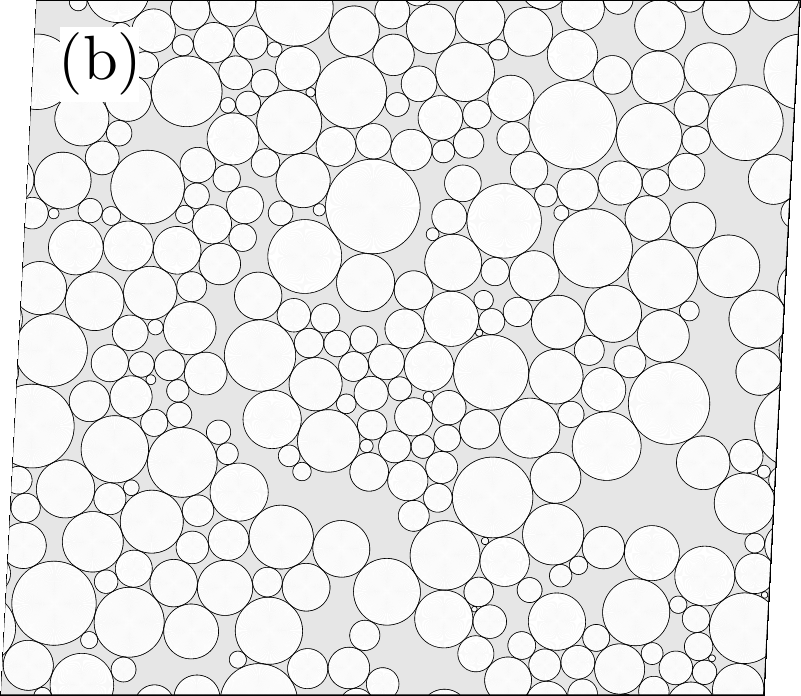}
		
		\vspace{1em}
		\includegraphics[width=0.7\columnwidth]{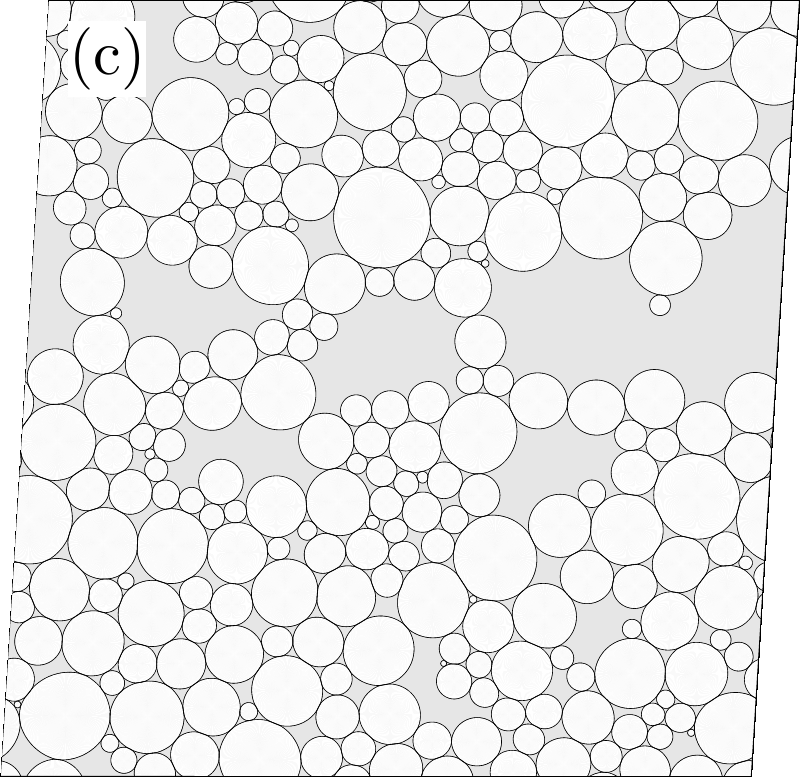}
		
		\caption{Simulated structure of $256$-bubble periodic foams at $\phi = 25\%$, for (a) repulsive $\Pi_\text{D}$, (b) $\theta = 0$, and (c) $\theta = 10^\circ$. The deformation of the domains is due to stress relaxation. Runs at $\theta = 10^\circ$ differ among themselves in the cleanliness of flocculation (i.e.\ whether a single tear forms). \label{fig:foam-struct-25p}}
	\end{figure}

	In Fig.~\ref{fig:foam-struct-25p}, we give the foam structure at $\phi = 25\%$ for several $\theta$. From this, it is clear that $\Pi_\text{O} \approx 0$ corresponds to a range of structures. For repulsive $\Pi_\text{D}$, the osmotic pressure is zero because the bubbles are not touching. However, when there is bubble attraction the foam flocculates, as found by \citeauthor{2018-cox} \cite{2018-cox} in simulated foams, and as is well known in emulsions~\cite{1993-bibette}. A clustered structure with many contacting bubbles is energetically favourable, due to the decrease of interface tension within films [recalling Eq.~\eqref{eqn:contact-angle}] \cite{1983-princen,2018-cox}. The presence of bubble attraction, and hence flocculation, for $\theta = 0$ is an artefact of our finite film thickness, as discussed in Section~\ref{sub:met:struct}.
	
	We note, from Fig.~\ref{fig:osmotic-press-v-liquid-fraction}, that $\Pi_\text{O} < 0$ for a range of $\phi$ near $16\%$ at contact angle $\theta = 10^\circ$ (including for the data with $h_0 = 0$). This would correspond to the foam exerting suction on the confining piston mentioned above. As $\phi$ is further increased, $\Pi_\text{O}$ increases toward zero. We interpret this behaviour as arising from the energy cost for a bubble to lose a neighbour at high $\theta$. Bubbles therefore experience some extension before they detach, and this regime, in which the foam is not fully flocculated, has $\Pi_\text{O} < 0$. Our interpretation is supported by the observation that, when $\phi$ is decreased again from $25\%$, $\Pi_\text{O}$ remains nonnegative (not shown). This hysteresis is in contrast to the behaviour for repulsive $\Pi_\text{D}$ \cite{1995-hutzler-b}.
	
\begin{figure}
		\includegraphics[width=1.0\columnwidth]{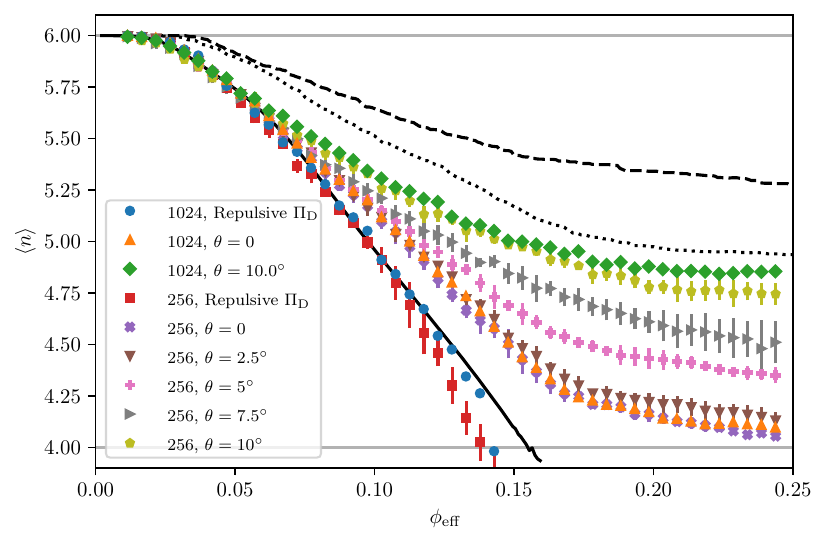}
		\caption{Mean number of neighbours $\langle n \rangle$ versus effective liquid fraction (defined in the text), for foams with various degrees of bubble attraction. Data from single runs are plotted for $1024$-bubble foams, while the data for $256$ bubbles is averaged over five runs (the error bars give the sample standard deviation). We compare with {\sc PLAT} data at $\theta = 0$ and $h_0 = 0$ (solid curve) \cite{2017-winkelmann,2021-jing}, along with $h_0 = 0$ simulations at $\theta = 10.8^\circ$ (dashed curve) and $\theta = 5.1^\circ$ (dotted curve) using the approach of \citeauthor{2021-jing} \cite{2021-jing} and \citeauthor{2018-cox} \cite{2018-cox} (the same runs as in Fig.~\ref{fig:osmotic-press-v-liquid-fraction}). \label{fig:mean-neigh-num-v-liq-frac}}
	\end{figure}

	Next, in Fig.~\ref{fig:mean-neigh-num-v-liq-frac}, we plot the mean number of bubble neighbours $\langle n \rangle$. We measure $n$ for a bubble $\text{B}$ by counting the distinct bubbles from whose interfaces the vertices of $\text{B}$ experience a positive (i.e.\ repulsive) $\Pi_\text{D}$. The behaviour of $\langle n \rangle$ has been studied in detail for $\theta > 0$~\cite{2021-jing,2021-feng,2023-jing}, and by \citeauthor{2017-winkelmann} \cite{2017-winkelmann} and \citeauthor{2019-boromand} \cite{2019-boromand} without bubble attraction (the latter using a comparable model to ours). Unlike prior studies, to our knowledge, we have used the same model for both cases.
	
	The PLAT simulation data \cite{2021-jing,2017-winkelmann} is in good agreement for smaller $\phi_\text{eff}$. Our deviation at larger values may be due to our finite film thickness. In Fig.~\ref{fig:mean-neigh-num-v-liq-frac}, we also compare with simulations using the same model as \citeauthor{2021-jing} \cite{2021-jing} for $\theta > 0$. The disagreement is much larger (surprisingly, given the consistency in Fig.~\ref{fig:osmotic-press-v-liquid-fraction}), although the qualitative behaviour \cite{2021-feng} is the same. The approximate plateaus in $\langle n \rangle$ are considerably lower in our simulations, and the failure of the decoration theorem ($\langle n \rangle < 6$ \cite{1990-bolton}) occurs at smaller $\phi_\text{eff}$. Our deviations from \citeauthor{2021-jing} \cite{2021-jing} are slightly smaller, but still large, and we find less of a difference than them between zero and slight bubble attraction.
	
	The differences may be due to the much lower polydispersity (defined in Sec.~\ref{sub:met:generate}) in the compared simulations: $\sigma = 0.008$ ($\theta = 10.8^\circ$) and $\sigma = 0.041$ ($\theta = 5.1^\circ$). We were able to reproduce a similar dependence of $\langle n \rangle$ to \citeauthor{2021-jing} \cite{2021-jing} in a $256$-bubble foam at $\theta = 10^\circ$, when the bubble areas were sampled from a triangular distribution with the same (very small) polydispersity that they used. A slight $\sigma$ dependence in $\langle n \rangle$ was observed by \citeauthor{2023-jing} \cite{2023-jing}.
	
	Care is needed when increasing $\phi$ in our simulations with bubble attraction, due to the fragility of the bubble clusters, which can easily be broken apart if $\Delta \phi$ is too large. This would decrease $\langle n \rangle$, although our convergence tests imply that this is not the cause of our differences from prior data.
	
	\subsection{Bubble properties \label{sub:coa:bubbles}}
	
	We now turn to the properties of individual bubbles. As noted in Sec.~\ref{sec:intro}, the growth rate of a bubble is determined partly by its pressure $p$ and the length $L$ of its perimeter along which it is in contact with its neighbours (i.e.\ its film length) \cite{1978-lemlich,2013-roth}. Therefore, we focus on these properties, before considering the growth rates themselves in Sec.~\ref{sub:coa:growth-rates}.
	
	\subsubsection{Bubble pressures \label{ssb:coa:pressure}}
	
\begin{figure}
		\includegraphics[width=1.0\columnwidth]{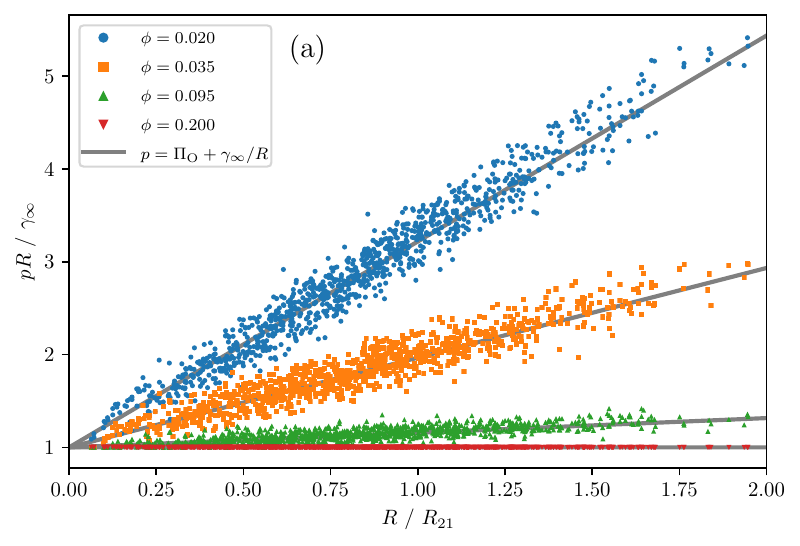}
		\includegraphics[width=1.0\columnwidth]{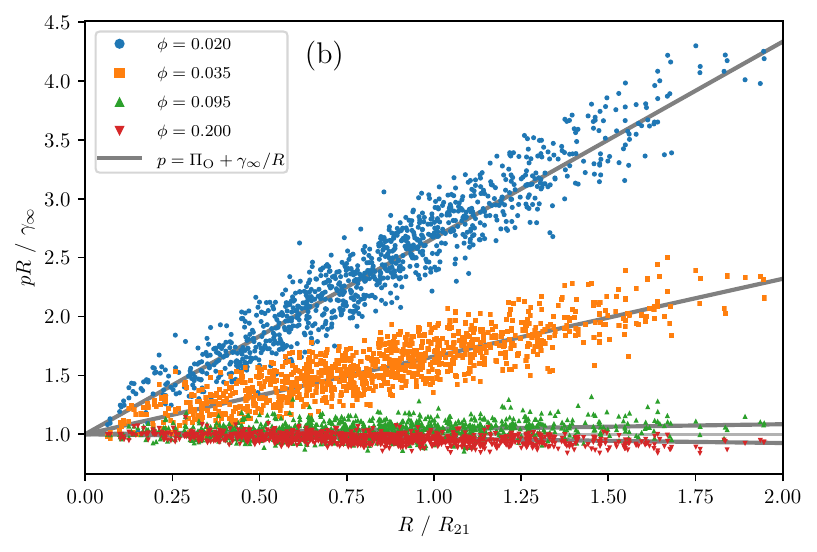}
		\caption{Scaled pressure versus effective radius, for individual bubbles in $1024$-bubble foams at various liquid fractions. Data for repulsive $\Pi_\text{D}$ (a) and $\theta = 10^\circ$ (b) is shown, and compared with Eq.~\eqref{eqn:eff-yl} using the osmotic pressure measured in the simulations. \label{fig:pressure-v-radius}}
	\end{figure}

\begin{figure}
		\includegraphics[width=1.0\columnwidth]{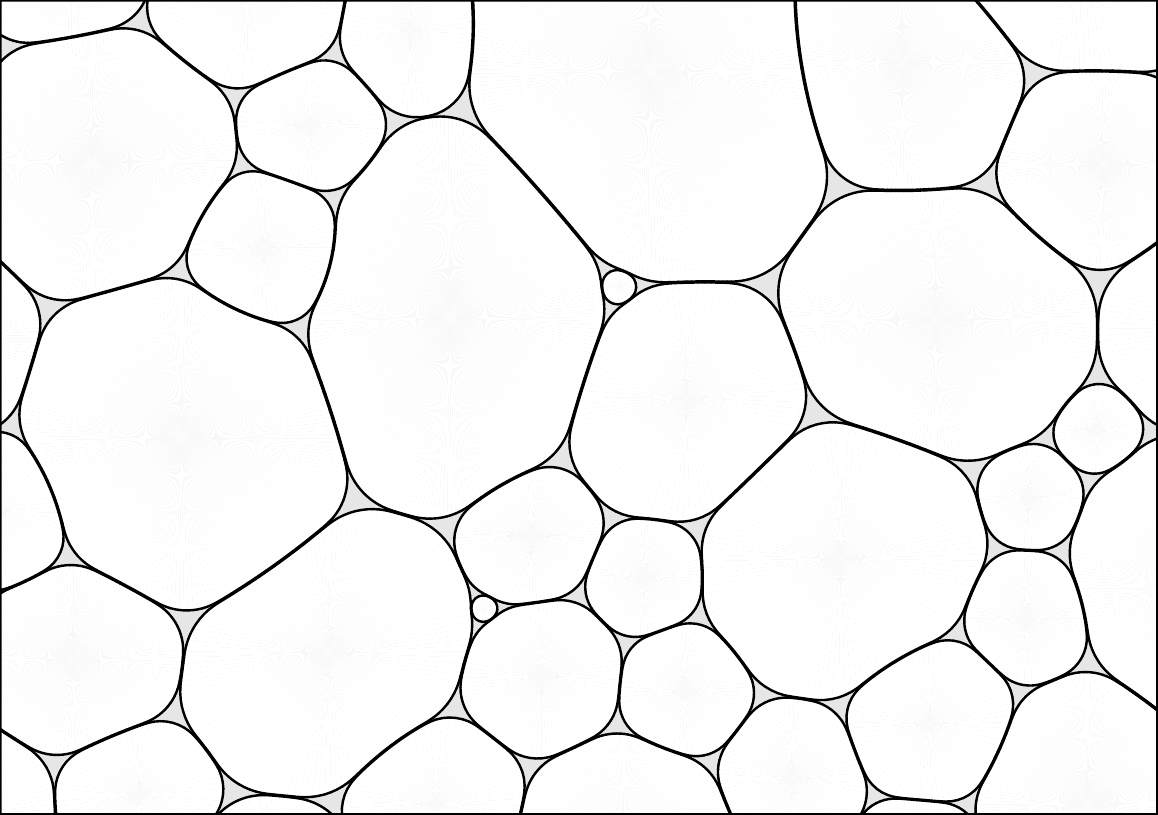}
		\caption{Illustration of the observation that larger bubbles are more deformed in a wet foam \cite{1991-bolton}, using a simulation with $\phi = 3.5\%$ and repulsive $\Pi_\text{D}$. The smallest bubbles are almost circular. \label{fig:rattler}}
	\end{figure}

	In Fig.~\ref{fig:pressure-v-radius}, we plot the scaled pressures of the bubbles against their effective radii $R$. Let $r$ be the radius of curvature of a bubble's Plateau borders, given via the Young-Laplace law $p = \gamma_\infty / r$ \cite{1999-weaire}. Then the plotted property is $p R / \gamma_\infty = R / r$. From the figure, this ratio approaches unity as $R \to 0$; i.e.\ the smallest bubbles are circular and undeformed. These are analogous to ``rattlers'' which exist in the Plateau borders of a polydisperse foam \cite{2018-khakalo,2023-galvani}. At smaller liquid fractions, $R / r$ increases with $R$, so larger bubbles experience relatively more deformation \cite{1991-bolton}. This is illustrated in Fig.~\ref{fig:rattler}. At higher liquid fractions, $R / r \approx 1$ for all $R$, as expected for repulsive $\Pi_\text{D}$, because all bubbles are approximately circular near and beyond the jamming transition \cite{1995-durian}. However, we see that $R / r \approx 1$ also holds in flocculated foams with bubble attraction (consistent with theory for ordered foams \cite{1979-princen}), although there is a greater spread in $R / r$ for large $\theta$ at all $\phi$.
	
	The relationship between $R$ and $p R$ appears linear to a good approximation. To interpret this, we use the approach of \citeauthor{2021-hohler} \cite{2021-hohler}. They generalized to arbitrary polydispersity an established relation~\cite{1988-princen} between the osmotic and capillary pressures in three-dimensional monodisperse foams. The same proof in two dimensions gives
	\begin{align}
		\Pi_\text{C} \approx \frac{\Pi_\text{O}}{1 - \phi} + \frac{\gamma_\infty}{R_{2 1}} .
			\label{eqn:cap-osm-rel}
	\end{align}
	We adapt their argument, which they apply over the entire foam, to a single bubble of effective radius $R$, and make additional simplifying assumptions. Let $\bar{\bm{\tau}}$ be the stress tensor spatially averaged over the considered bubble, including its interface.  We note that averaged stresses over individual bubbles in a dry foam were considered by \citeauthor{2012-kraynik} \cite{2012-kraynik}. Also let $\bar{p} = -\tfrac{1}{2} \Tr \bar{\bm{\tau}}$ be the bubble's reduced pressure (defined analogously to $\Pi_\text{O}$ \cite{2021-hohler}), which differs from its gas pressure $p$. Let $\Gamma$ be the bubble's interface. Using \citeauthor{1970-batchelor}'s expression for the averaged stress \cite{1970-batchelor,2021-hohler}, we obtain
	\begin{align}
		\bar{p} = p - \frac{1}{2 \pi R^2} \, \oint_\Gamma d l \ \gamma(h) ,
			\label{eqn:avg-press}
	\end{align}
	where $\gamma$ is the surface tension and $h$ the film thickness at a point on $\Gamma$. Next, we evaluate the integral [using that $\gamma(h) \approx \gamma_\infty$ in the films, from Eq.~\eqref{eqn:contact-angle} and that we have $\theta \lesssim 10^\circ$] and approximate the bubble perimeter as that of a circle with the same area (the error here is around $6\%$ for a simulated dry foam \cite{2012-kraynik}, and expected to be smaller for larger $\phi$). Hence, we obtain
	\begin{align}
		\bar{p} \approx p - \frac{\gamma_\infty}{R} . \label{eqn:avg-press-simp}
	\end{align}
	As a first approximation, we take the reduced pressure $\bar{p}$ to be the same for all bubbles. Since $\bar{p}$ is related to the forces applied externally to the bubble \cite[p.\ 7]{1986-landau} (see Appendix), this corresponds to a mean-field assumption~\cite{1978-lemlich,2006-feitosa} that the environment of each bubble is comparable. Substituting Eq.~\eqref{eqn:avg-press-simp} into Eq.~\eqref{eqn:cap-osm-rel}, and using that $\Pi_\text{C}$ is the area-weighted sum of bubble pressures $p$~\cite{2021-hohler}, gives $\bar{p} = \Pi_\text{O} / (1 - \phi)$. Hence, for $\theta \lesssim 10^\circ$,
	\begin{align}
		p \approx \Pi_\text{O} + \frac{\gamma_\infty}{R} , \label{eqn:eff-yl}
	\end{align}
	omitting the factor of $1-\phi$, which we find is a good approximation for all considered $\phi$ in two dimensions (noting that $\Pi_\text{O}$ typically decreases as $\phi$ increases). This is the single-bubble analogue of the more rigorous Eq.~\eqref{eqn:cap-osm-rel}. Its agreement with the data, shown in Fig.~\ref{fig:pressure-v-radius}, indicates that bubbles approximately obey an effective Young-Laplace law, with a mean-field external pressure $\Pi_\text{O}$ (whose value is currently taken from the simulations, but not fitted). Agreement also exists when the bubble areas are taken directly from a Voronoi tessellation (not shown). We therefore suggest that the pressure environment of the bubbles, determined by their neighbours, is comparable throughout the foam, and does not have a strong systematic variation with the bubble size.
	
	We are not aware of prior discussions of Eq.~\eqref{eqn:eff-yl} for polydisperse wet foams in the literature, although a comparable exact result for dry foams (when approximated as having straight films) is known~\cite{2023-li}. $\Pi_\text{O}$ diverges in this limit, so the relation is between the radius of a bubble and its pressure relative to some zero. It would be interesting to see, through similar simulations, whether this type of simple approximation also holds in three-dimensional polydisperse wet foams. As shown in the next section, Eq.~\eqref{eqn:eff-yl} may be applied to develop models for other bubble properties.
	
	\subsubsection{Film lengths \label{ssb:coa:film-lengths}}
	
\begin{figure}
		\includegraphics[width=1.0\columnwidth]{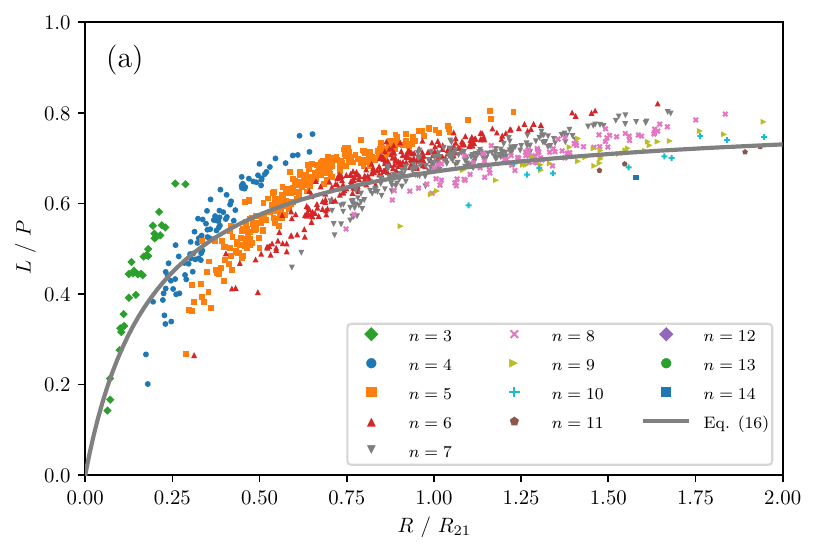}
		\includegraphics[width=1.0\columnwidth]{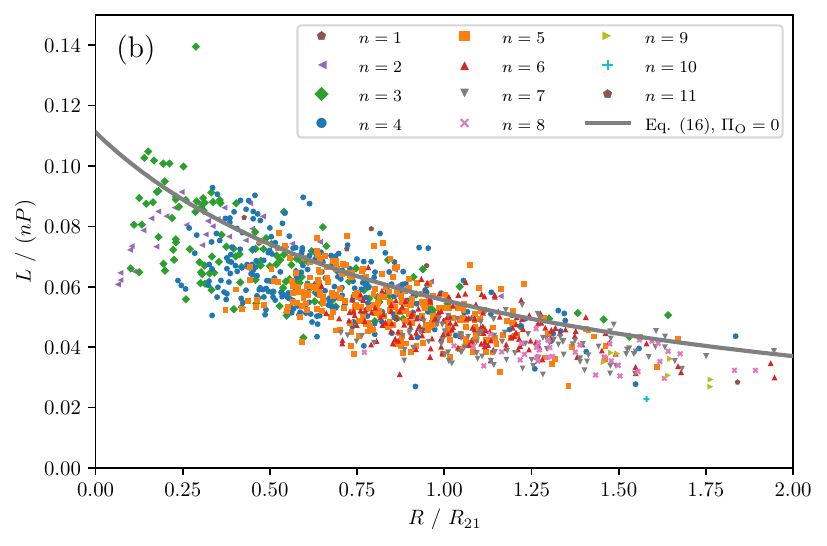}
		\caption{Ratio of film length to perimeter versus effective radius for individual bubbles (with $n$ neighbours) in a $1024$-bubble foam, at $\phi = 2\%$ with repulsive $\Pi_\text{D}$ (a), and at $\phi = 25\%$ with $\theta = 10^\circ$ (b). Comparison is made with Eq.~\eqref{eqn:film-length-approx}, with $\Pi_\text{O}$ taken from the simulations in (a), and assumed to be zero in (b) due to flocculation. The length per film $L / n$ is plotted in (b), since this is predicted directly by Eq.~\eqref{eqn:film-length-approx} for $\Pi_\text{O} = 0$. \label{fig:film-length-v-radius}}
	\end{figure}

	We now consider the length $L$ of the films adjoining each bubble in the simulated foams, which we recall is another determinant of growth rate. These are plotted in Fig.~\ref{fig:film-length-v-radius} for small and large $\phi$. We plot the ratio of $L$ to the bubble perimeter $P$; i.e.\ the proportion of bubble interface adjoining a film. Comparably to \citeauthor{1995-denkov} \cite{1995-denkov}, we measure $L$ in our simulations by approximating each film interface with a circular arc, and determining its intersection with (or closest approach to) circular arcs approximating the two adjoining Plateau-border interfaces. These interfaces are expected to approach circular arcs for high levels of mesh refinement, by the Young-Laplace law \cite{1999-weaire}.
	
	At $\phi = 2\%$, Fig.~\ref{fig:film-length-v-radius} shows that $L / P$ decreases with decreasing $R$, in agreement with the observation that smaller bubbles experience less deformation \cite{1991-bolton}. As $R$ increases, $L / P$ appears to saturate. No qualitative variation in this behaviour with $\theta$ was observed. At large $\phi$, where $L$ is smaller due to decreased bubble deformation, the film proportion depends strongly on $\theta$ due to flocculation, as shown in Sec.~\ref{sub:coa:bubbles-avg}.
	
	Certain features of $L / P$ may be captured by an analytic approximation, as for the bubble pressures. Let $\bar{R} \equiv R / R_{2 1}$, and $\bar{\Pi} \equiv \Pi_\text{O} R_{2 1} / \gamma_\infty$ \cite{1988-princen}. In the Appendix, we argue that
	\begin{align}
		\frac{L}{P} \approx 2 \, \frac{\bar{\Pi} \, \bar{R} + n \, \theta / \pi}{1 + \left(2 \bar{\Pi} + 1\right) \! \bar{R}} . \label{eqn:film-length-approx}
	\end{align}
	This is derived by relating the reduced bubble pressure $\bar{p}$ to the forces applied to the bubble's interface by its neighbours, and approximating this pressure by $\Pi_\text{O}$ as previously. We assume that $\theta \lesssim 10^\circ$, as considered in our simulations. The predictions of this equation are compared with the simulated values of $L$ in Fig.~\ref{fig:film-length-v-radius}.
	
	The general variation of $L$ with $R$ at low liquid fractions and for repulsive $\Pi_\text{D}$ is captured, although the spread at a given $R$ is not. This is attributed to the mean-field approach to neighbouring bubbles, while we expect the decoration theorem \cite{1991-bolton} to induce strong neighbour correlations for small $\phi$ (by constraining the film curvatures through Plateau's laws \cite{1999-weaire}). For foams with $\theta > 0$, an approximate relation between $n$ and $R$ in wet foams is needed to predict $L$ as a function of $R$ only. This is beyond our scope, and so we gauge Eq.~\eqref{eqn:film-length-approx} via $L / (n P)$ (the length per film) in flocculated foams for which $\Pi_\text{O} \approx 0$. Again, a broad agreement with the data is seen.
	
	An isotropic model \cite{2013-roth} might reproduce the spread in $L$ at small $\phi$, including the bands for different neighbour numbers $n$. However its reliance on the decoration theorem precludes applicability to moderate $\phi$, which is our regime of interest.
	
	The behaviour of $L$ appears more complex than that of $p$, described previously. Nevertheless, an approximate averaging of Eq.~\eqref{eqn:film-length-approx} describes the mean film lengths very well, as shown in Sec.~\ref{sub:coa:bubbles-avg}.
	
	\subsection{Bubble growth rates \label{sub:coa:growth-rates}}
	
	We now summarize the simulated bubble growth rates, recalling that these include gas transfer across films and through Plateau borders, as described in Sec.~\ref{sub:met:coarsening}. We describe how they change as $\phi$ is increased, and compare our simulations with prior models, including the relationship between a bubble's growth rate and its effective number of neighbours $n_\text{eff}$ \cite{2012-fortuna}.
	
	\subsubsection{Effects of liquid fraction \label{ssb:coa:liq-frac}}
	
\begin{figure}
		\includegraphics[width=1.0\columnwidth]{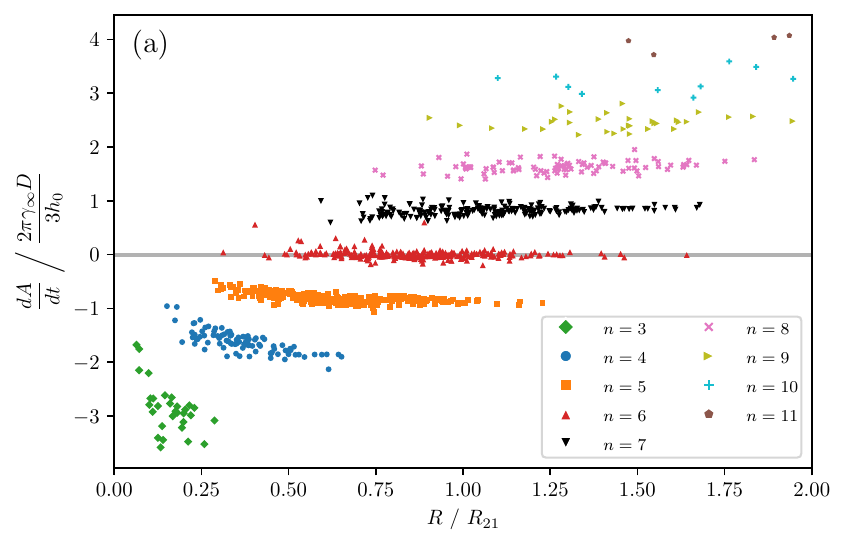}
		\includegraphics[width=1.0\columnwidth]{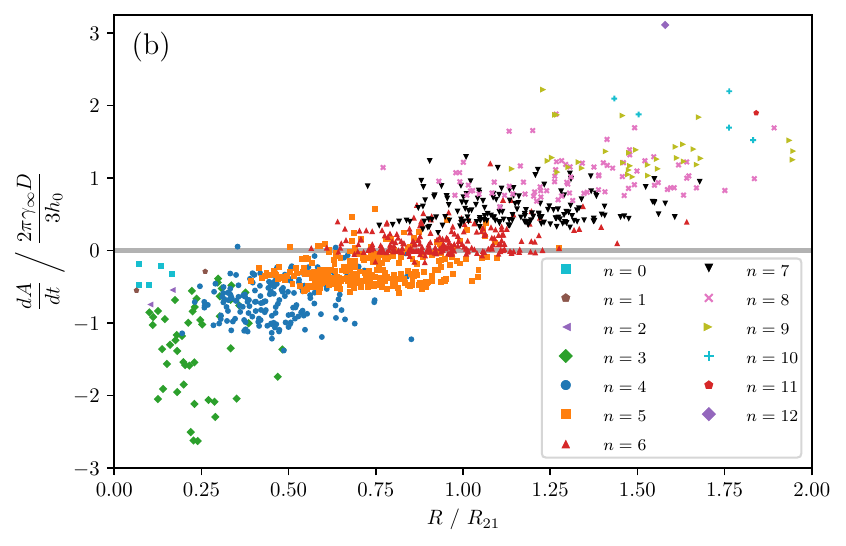}
		\includegraphics[width=1.0\columnwidth]{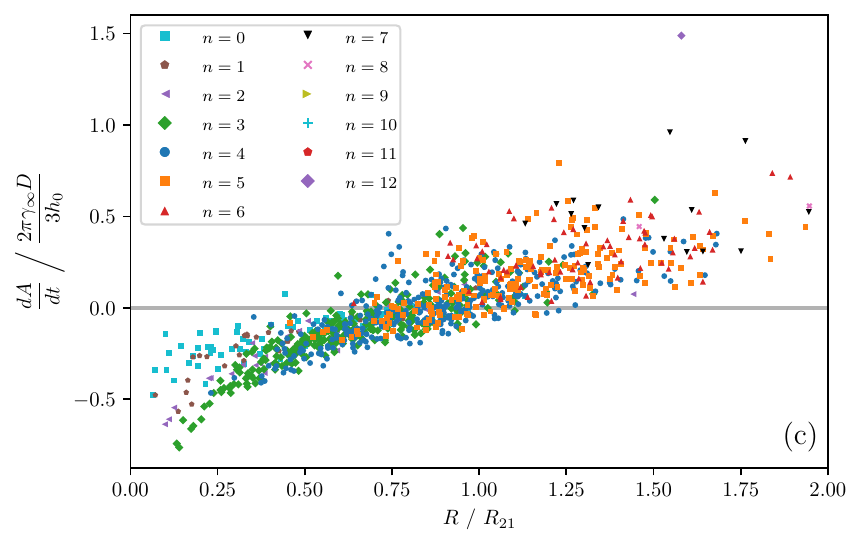}
		\caption{Individual bubbles growth rates versus effective radius in a particular $1024$-bubble foam at different liquid fractions $\phi$, with repulsive $\Pi_\text{D}$. The number of neighbours $n$ of each bubble is shown. In (a), $\phi = 2\%$; in (b), $\phi = 6.5\%$; and in (c), $\phi = 15.5\%$. The growth rates are scaled so that the rates from von Neumann's law, Eq.~\eqref{eqn:von-neumann}, are the integers. A bubble with $n = 14$ lies outside the plotted domain in (a), to better show the level structure. \label{fig:growth-rates}}
	\end{figure}

	At $\phi = 0$, which is inaccessible to our simulations, the growth rate of a bubble with $n$ neighbours obeys von Neumann's law \cite{1952-von-neumann}:
	\begin{align}
		\dot{A} = \frac{2 \pi \gamma_\infty D}{3 h_0} \left(n - 6\right) .
			\label{eqn:von-neumann}
	\end{align}
	The prefactor of $n - 6$, which is constant for a particular foam, is derived, for example, by \citeauthor{2013-cantat} \cite[p.\ 109]{2013-cantat}. When $\phi$ is increased to small values, such that almost all Plateau borders still connect exactly three films (i.e.\ the decoration theorem is approximately satisfied \cite{1991-bolton}), the growth rates remain predominantly determined by $n$, although the reduction of gas flow by the Plateau borders induces a dependence upon bubble size and shape \cite{1991-bolton,2013-roth}. In Fig.~\ref{fig:growth-rates}(a), we plot the bubble growth rates at $\phi = 2\%$, noting that the qualitative behaviour is in agreement with experiments and theory for quasi-two-dimensional foams --- in particular, for fixed $n$, the absolute growth rates tend to increase slightly with increasing bubble size \cite{2013-roth}. This implies that larger bubbles experience less border blocking \cite{1991-bolton}, again consistent with their greater observed deformation. As discussed later, the absolute growth rates of the smallest bubbles are enhanced by their thinner films in our simulations.
	
	When $\phi$ is increased further, the Plateau borders grow, and the degree of border blocking increases \cite{2013-roth}. The primary effect of this is to reduce the absolute growth rates, as shown in Fig.~\ref{fig:growth-rates}(b). However, the decoration theorem also fails for more bubbles as they detach from their neighbours with increasing $\phi$. The geometric constraints from Plateau's laws \cite{1999-weaire} are relaxed for such bubbles, resulting in growth rates that lie between the levels hitherto occupied.
	
	Eventually, $\phi$ is high enough that most of the bubbles adjoin at least one Plateau border which connects more than three films. As shown in Fig.~\ref{fig:growth-rates}(c), a level structure is no longer discernible.
	
	The absolute growth rates continue to decrease with increasing $\phi$, except in the presence of adhesion, where they eventually plateau due to flocculation. This is considered further in Sec.~\ref{sub:coa:bubbles-avg}. Due to the inclusion of gas flow through the Plateau borders in our model, the growth rates do not shrink to zero at the jamming transition $\phi_\text{c} \approx 16\%$ for repulsive $\Pi_\text{D}$. Instead they continue to decrease in magnitude as the bubbles move apart. We expect that our approximation for the growth rates, from Sec.~\ref{sub:met:coarsening}, declines in accuracy as $\phi$ increases beyond this transition, as the gas dissolved in the bulk liquid becomes the main source and sink for bubbles, rather than pairwise transfer between adjacent bubbles \cite{2012-fortuna}. However, as justified below, we believe that the approximation remains valid for flocculated foams, since bubbles remain in contact.
	
	\subsubsection{Effective neighbour number \label{ssb:coa:eff-neigh-num}}
	
\begin{figure}
		\includegraphics[width=1.0\columnwidth]{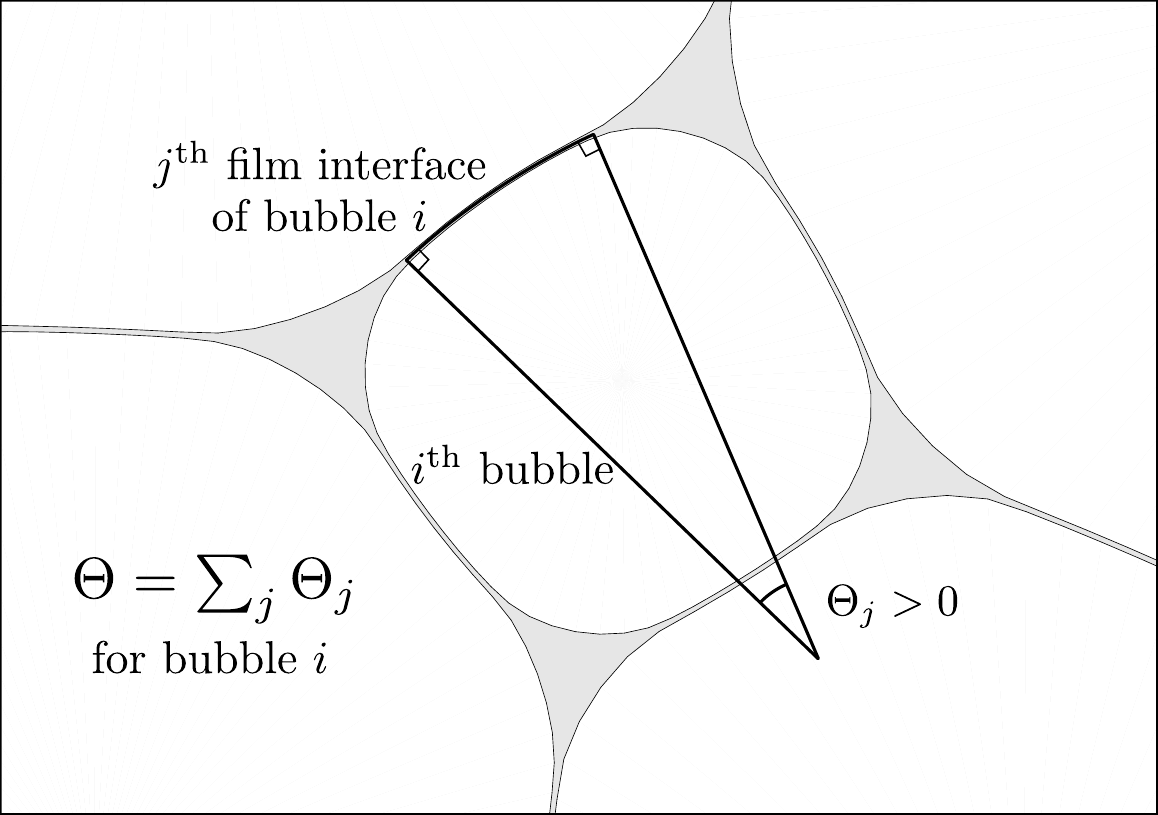}
		\caption{Schematic of the definition of the angle $\Theta$ \cite{2012-fortuna}, described in the text. \label{fig:eff-neigh-num-def}}
	\end{figure}

	We now compare our simulated growth rates with results from the literature. \citeauthor{2012-fortuna} \cite{2012-fortuna} introduced the effective number of neighbours $n_\text{eff}$ of a bubble in a two-dimensional wet foam. Let $\Theta$ be the angle turned by the bubble interface's tangent as its film interfaces are traversed, illustrated in Fig.~\ref{fig:eff-neigh-num-def}. Contributions to $\Theta$ from a convex interface, as in this figure, are positive. Then \cite{2012-fortuna}
	\begin{align}
		n_\text{eff} = 6 - 3 \Theta / \pi . \label{eqn:eff-neigh-num}
	\end{align}
	This is equivalent to the bubble's actual number of neighbours $n$ for $\phi = 0$, and is equal to $6$ in a wet foam when $n = 0$ \cite{2012-fortuna}. If $n_\text{eff} < 6$, then Eq.~\eqref{eqn:eff-neigh-num} implies that the film interfaces of the bubble tend to be convex, and so its pressure tends to be larger than those of its neighbours; i.e.\ it is smaller than average, in a loose sense, from Eq.~\eqref{eqn:eff-yl}. The converse is true for $n_\text{eff} > 6$. In fact, by repeating the derivation \cite{1952-von-neumann} of von Neumann's law in a wet foam, it can be shown that the border-blocking growth rate of a bubble (i.e.\ including only gas flow through the films) is exactly
	\begin{align}
	\dot{A}_\text{BB} = \frac{2 \pi \gamma_\infty D}{3 h_0} \left(n_\text{eff}
	- 6\right) . \label{eqn:growth-rate-border-block}
	\end{align}
	This is von Neumann's law with $n$ replaced by $n_\text{eff}$, and was derived in a slightly different form by \citeauthor{1991-bolton} \cite{1991-bolton}. \citeauthor{2012-fortuna} \cite{2012-fortuna} validate this result, in conjunction with an approximation for the Plateau border gas flow, using Potts model simulations.
	
	We measure $\Theta$, and hence $n_\text{eff}$, by noting that the contribution to $\Theta$ from a bubble's $j^\text{th}$ film is $l_j / r_j$~\cite{2012-fortuna}, where $l_j$ is the film's length, and $r_j$ is its radius of curvature (which is positive if the film is convex, and obtained via the arithmetic mean of the two interface curvatures, from the augmented Young-Laplace equation~\cite{1985-kralchevsky-b,1995-denkov}).
	
\begin{figure}
		\includegraphics[width=1.0\columnwidth]{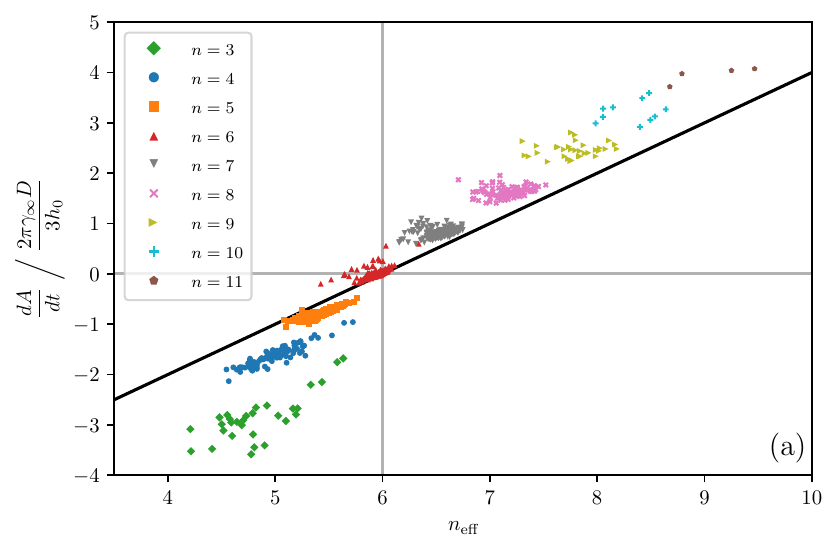}
		\includegraphics[width=1.0\columnwidth]{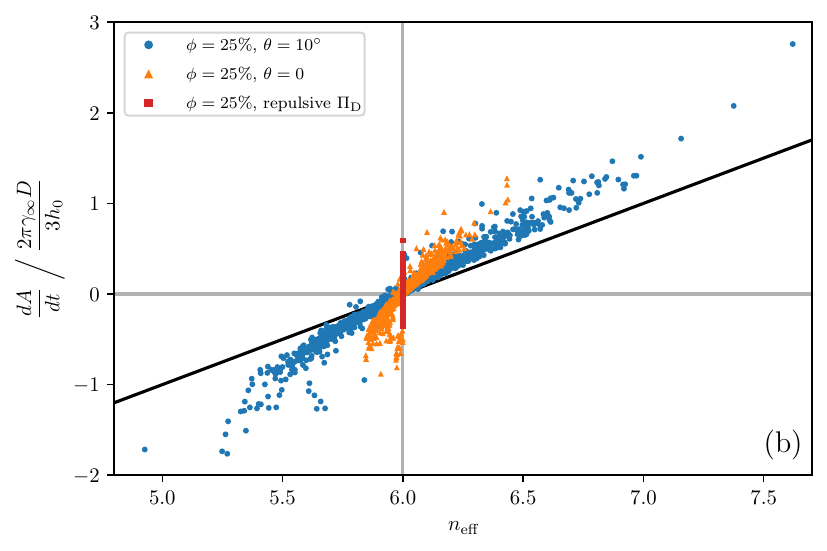}
		\caption{Bubble growth rate versus $n_\text{eff}$ in a $1024$-bubble foam (a) at $\phi = 2\%$ for repulsive $\Pi_\text{D}$, and (b) at $\phi = 25\%$ for different $\theta$. A large bubble with $n = 14$ neighbours is omitted from (a), for clarity. The black lines are Eq.~\eqref{eqn:growth-rate-border-block} for the set film thickness $h_0$. \label{fig:growth-rate-v-n-eff}}
	\end{figure}

	Fig.~\ref{fig:growth-rate-v-n-eff} shows our bubble growth rates $\dot{A}$ plotted against $n_\text{eff}$ (including contributions from gas flow through the Plateau borders). The two quantities are closely related at all $\phi$ and $\theta$ we have investigated, provided that the bubbles have neighbours (including in flocculated foams). Otherwise $n_\text{eff} = 6$ identically, while the growth rates still vary.
	
	This close relationship between $\dot{A}$ and $n_\text{eff}$ (with the sign of $\dot{A}$ being well determined by the sign of $n_\text{eff} - 6$) was not apparent to us from the simulations performed by \citeauthor{2012-fortuna} \cite{2012-fortuna}, likely due to the fact that the Potts model does not accurately capture bubble geometry. In principle, deviations from Eq.~\eqref{eqn:growth-rate-border-block} give the gas transfer through Plateau borders, although the following caveats should be considered.
	
	A separate group of bubbles may be observed in Fig.~\ref{fig:growth-rate-v-n-eff} at $\phi = 25\%$ and $\theta = 0$, with $n_\text{eff} < 6$ and growth rates considerably smaller than other bubbles at the same $n_\text{eff}$ (close to the vertical line). These are interpreted to result from difficulties in defining and measuring film lengths and radii of curvature in extremely short films with a film thickness comparable to their length. The group is smaller at higher mesh refinement, but is not practical to eliminate.
	
\begin{figure}
		\includegraphics[width=1.0\columnwidth]{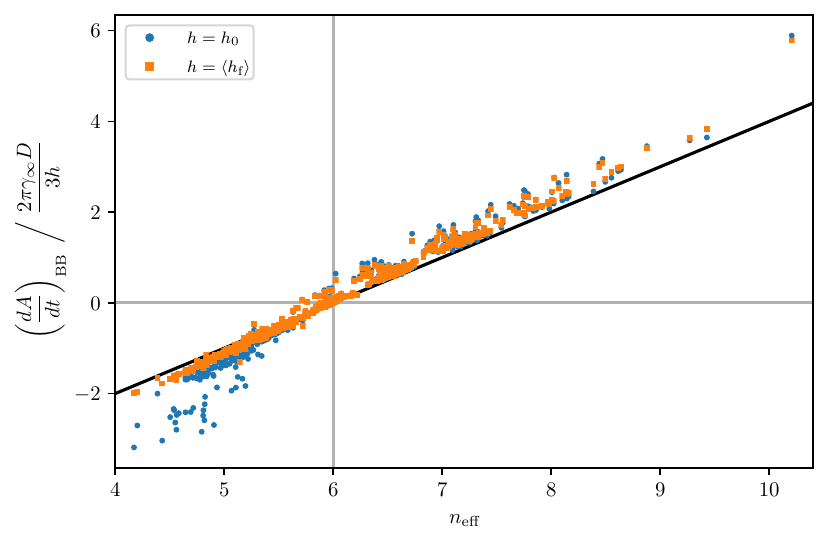}
		\caption{Simulated border-blocking bubble growth rates in a $1024$-bubble foam with repulsive $\Pi_\text{D}$ at $\phi = 2\%$, compared with their scalings to approximately account for film thickness variations between bubbles. $\langle h_\text{f} \rangle$ is a bubble's mean film thickness --- one representative thickness is measured near the centre of each of its films, and the mean of these thicknesses is calculated. The black line is Eq.~\eqref{eqn:growth-rate-border-block} with $h_0$ replaced by $h$. \label{fig:growth-rate-bb-v-n-eff}}
	\end{figure}

	We previously mentioned film thickness variations due to our form for $\Pi_\text{D}$. The effect of these on the bubble growth rates is shown in Fig.~\ref{fig:growth-rate-bb-v-n-eff}, where we plot our simulated border-blocking growth rates (given by summing the product of film length and pressure difference of the bubble from its neighbour, divided by film thickness, over each of a bubble's films~\cite{1991-bolton,2013-cantat}) against Eq.~\eqref{eqn:growth-rate-border-block}, for which the thickness of each film is the set value $h_0$. Hence, large deviations are mainly restricted to small bubbles. We approximately correct for the film thickness variations by scaling the growth rates with the measured mean film thickness $\langle h_\text{f} \rangle$ (see caption of Fig.~\ref{fig:growth-rate-bb-v-n-eff}), reducing the differences for small bubbles. This simple scaling cannot be done for the total growth rates, since the dependence on film thickness of Plateau border transfer is different, as seen in the next section.
	
	While discrepancies are still present, part of which we expect to come from thickness variations between the individual films of each bubble (agreement is not substantially improved at higher mesh refinement), we are consistent with Eq.~\eqref{eqn:growth-rate-border-block} (recalling that we calculate our growth rates and $n_\text{eff}$ using different techniques). This supports the use of our coarsening approximation (Sec.~\ref{sub:met:coarsening}) to obtain the gas transfer rates through films.
	
	At $\phi = 25\%$, the growth rates are also affected by the fact that the foam-wide mean film thickness is larger than the set value $h_0$ (by a factor of approximately $1.59$ for $\theta = 0$). We interpret this by observing that the interfaces of short films between bubbles with equal pressure retain curvature over their whole lengths in our simulations --- the augmented Young-Laplace equation~\cite{1985-kralchevsky-b,1995-denkov} then requires a weaker disjoining pressure than for zero curvature, and hence a thicker film.
	
	\subsubsection{Gas transfer through Plateau borders \label{ssb:coa:crossing}}
	
\begin{figure}
		\includegraphics[width=1.0\columnwidth]{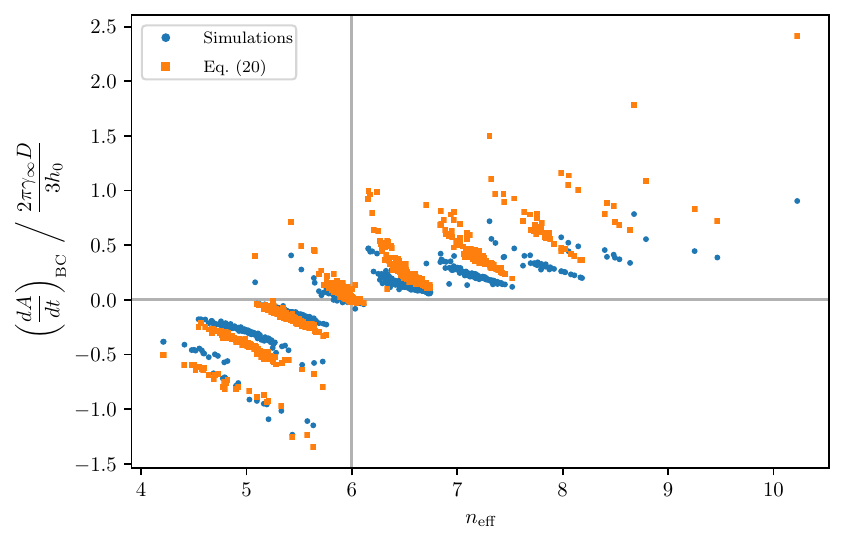}
		\caption{Border-crossing growth rate, defined as the difference between the total and border-blocking rates, versus $n_\text{eff}$ in a $1024$ bubble foam at $\phi = 2\%$ with repulsive $\Pi_\text{D}$. The simulations are compared with Eq.~\eqref{eqn:growth-rate-border-cross}. \label{fig:cross-rate-v-n-eff}}
	\end{figure}

	\citeauthor{2017-schimming} \cite{2017-schimming} derive a bubble growth law in two-dimensional wet foams at small $\phi$ (satisfying the decoration theorem), which accounts for gas flow through the Plateau borders. It is assumed that these borders are symmetric, and that $\theta = 0$. Let $r$ be the radius of curvature of a bubble's Plateau borders, let $r_i$ be the signed radius of curvature of its $i^\text{th}$ film (measured as in the previous section), and define the circularity $C = (R / n) \, \sum_{i = 1}^n 1/r_i$ \cite{2013-roth}. Then the contribution to the bubble's growth rate from Plateau border transfer (which they term border crossing) is \cite{2017-schimming} (corrected, we believe, by a factor of $\sqrt{3}$)
	\begin{align}
		\dot{A}_\text{BC} \approx -2 \pi \gamma_\infty D \ \frac{n C}{R}
			\, \sqrt{\frac{r}{h_0}} . \label{eqn:growth-rate-border-cross}
	\end{align}
	We plot a comparison between this approximation and our results for a repulsive disjoining pressure at $\phi = 2\%$ in Fig.~\ref{fig:cross-rate-v-n-eff}. Most of the difference is believed to come from the transition regions at the ends of the films~\cite{1985-kralchevsky-b} --- the approach we use to measure the film lengths tends to include part of these regions [which are omitted in the derivation of Eq.~\eqref{eqn:growth-rate-border-cross}], thus overestimating the border-blocking growth rates. We note that the definition of film length is ambiguous for finite film thickness, due to these transition regions. Agreement is not improved at higher mesh refinement (and is worsened for the smallest bubbles, in the cluster with smallest $\dot{A}_\text{BC}$).
	
	We recall that our selected $h_0$ is large compared to real foams. This is expected to exaggerate the border-crossing rates relative to the gas flow through the films in our simulations \cite{2017-schimming}.
	
\begin{figure}
		\includegraphics[width=1.0\columnwidth]{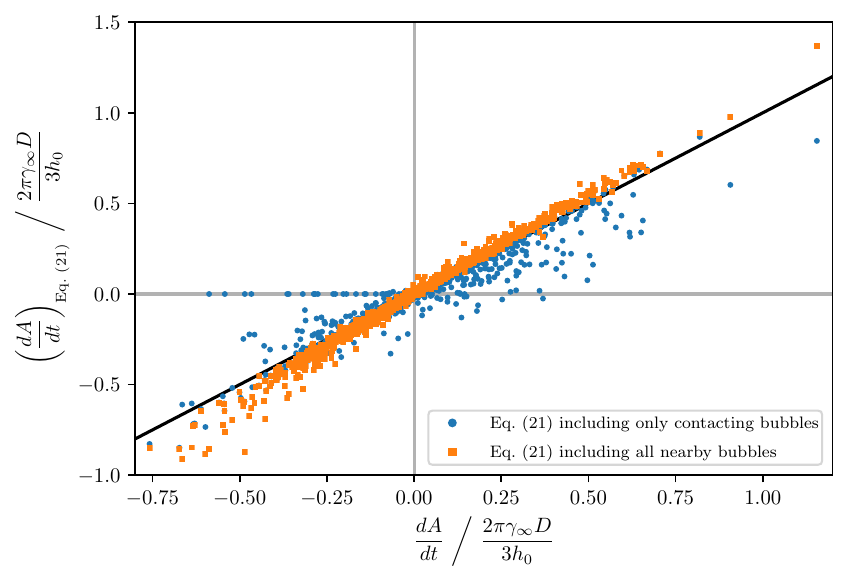}
		\caption{Total bubble growth rates plotted against Eq.~\eqref{eqn:growth-rate-jamming} for a $1024$-bubble foam at $\phi = 16\%$ (set in one step, rather than increased gradually) with repulsive $\Pi_\text{D}$. We do not expect the means of setting $\phi$ to have a large effect for this $\Pi_\text{D}$. Contacting bubbles are those which exert $\Pi_\text{D} > 0$ on a vertex of the considered bubble, and nearby bubbles are those counted as neighbours by the vertex neighbour algorithm (see Sec.~\ref{sub:met:struct} and Fig.~\ref{fig:bubble-neighbours}). The black line has unit gradient. \label{fig:jamming-rate-v-approx}}
	\end{figure}

	Near to the jamming transition, at which the films vanish, the bubble growth rate due to one neighbour was given by \citeauthor{2017-schimming} \cite{2017-schimming} for $\theta = 0$. Let $R_i$ be the effective radius of the bubble's $i^\text{th}$ neighbour, and $h_i$ the minimum distance between their interfaces. By adding contributions from each neighbour (neglecting the effect of other neighbours on each contribution, which should tend to reduce the absolute growth rates \cite{2002-evilevitch}),
	\begin{align}
		\dot{A} \approx \pi \gamma_\infty D \, \sum_{i = 1}^n \sqrt{\frac{2}{h_i}}
			\, \frac{R - R_i}{\sqrt{R R_i (R + R_i)}} . \label{eqn:growth-rate-jamming}
	\end{align}
	This is compared with our simulations in Fig.~\ref{fig:jamming-rate-v-approx}, which shows good agreement, provided that contributions from all nearby bubbles are included, not just those in contact.
	
	While the issue of accounting for long-range gas transfer in simulations \cite{2022-hohler} remains unresolved (for which our approximation described in Sec.~\ref{sub:met:coarsening} is likely poor), and may be particularly important due to our thick films, we observe here that the two different approaches to the bubble gas flow rates are consistent. Eq.~\eqref{eqn:growth-rate-jamming} uses an approximation for small $h_i$, although we find the exact prediction \cite{2017-schimming} (which gives larger absolute growth rates) differs by less than $10\%$ for $h_1 \leq 2 R$ when $R_1 \leq 10 \, R$ and $n = 1$. We expect that most of the difference between the two datasets plotted in Fig.~\ref{fig:jamming-rate-v-approx} is due to bubbles which are close to being in contact.

	The use of our growth rate approximation (Sec.~\ref{sub:met:coarsening}) for flocculated foams, in which bubbles retain close contacts (with short films) as $\phi$ is increased, is thereby supported, since the approximation is consistent with predictions of (short-range) gas transfer occurring wholly through the bulk liquid.
	
	\subsection{Statistics of bubble properties \label{sub:coa:bubbles-avg}}
	
	We now give averaged properties of the bubbles in the simulated foams, in order to quantify some of our observations regarding their variation with $\phi$ and $\theta$.
	
	\subsubsection{Film lengths \label{ssb:coa:length-stats}}
	
	First, we consider the averaged bubble film lengths $L$. Recalling our approximation, Eq.~\eqref{eqn:film-length-approx}, we may estimate its prediction of the mean ratio of film length to perimeter, for $\theta \lesssim 10^\circ$, as
	\begin{align}
		\left\langle \frac{L}{P} \right\rangle \approx \frac{\bar{\Pi}
			+ \langle n \rangle \, \theta / \pi}{1 + \bar{\Pi}} \label{eqn:film-length-avg}
	\end{align}
	by setting $R = R_{2 1}$ and $n = \langle n \rangle$ on the right-hand side. This has an equivalent form for $\theta = 0$ to an existing approximation for the film areas in a monodisperse three-dimensional foam \cite{2021-hohler,2023-pasquet-b}.
	
\begin{figure}
		\includegraphics[width=1.0\columnwidth]{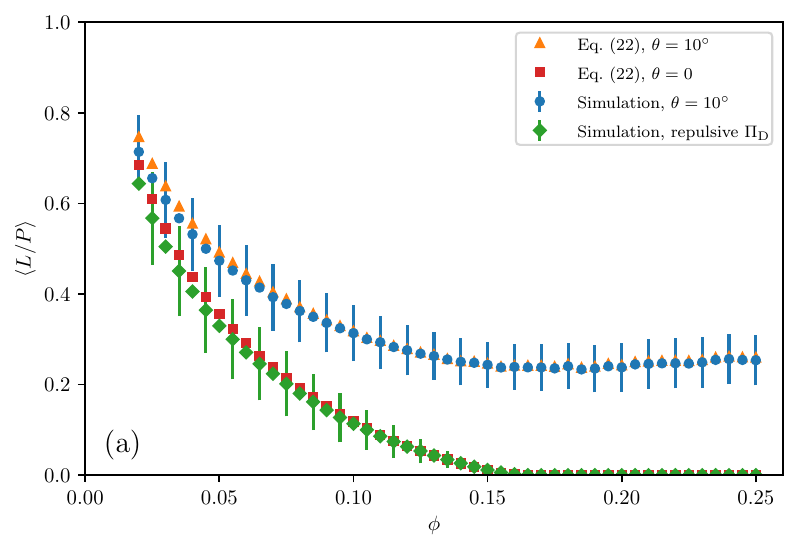}
		\includegraphics[width=1.0\columnwidth]{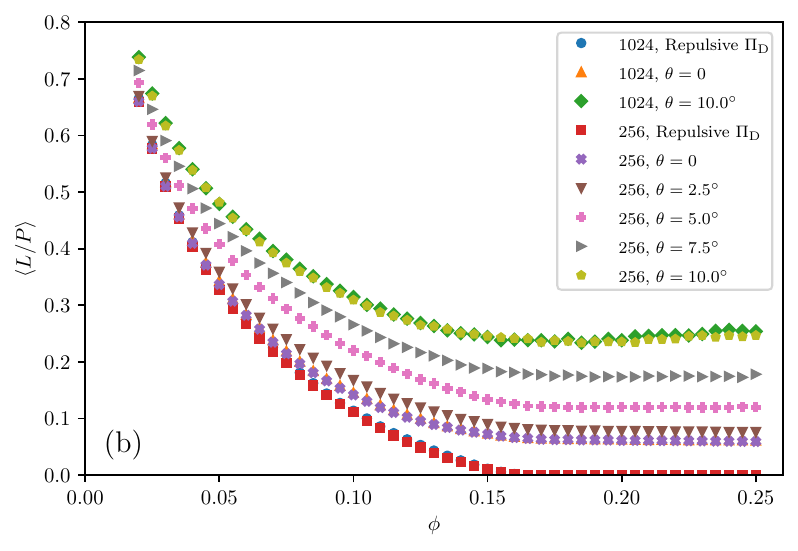}
		\caption{Mean ratio of film length to perimeter versus $\phi$, for repulsive $\Pi_\text{D}$ alongside various $\theta$. In (a), simulations with $1024$ bubbles are compared with Eq.~\eqref{eqn:film-length-avg}, and error bars give the standard deviation of the simulated $L / P$ distribution. In (b), single runs with $1024$ bubbles are shown alongside the mean of $5$ runs with $256$ bubbles. Error bars in (b), giving the sample standard deviation of $\langle L / P \rangle$ in the latter simulations, are omitted since they are close to the markers in size. \label{fig:film-length-v-liq-frac}}
	\end{figure}

	Equation~\eqref{eqn:film-length-avg} is compared with our simulations in Fig.~\ref{fig:film-length-v-liq-frac}, and we observe a very close agreement, including in flocculated foams. We note that our simulations with repulsive $\Pi_\text{D}$, rather than $\theta = 0$, should be compared with the predictions of Eq.~\eqref{eqn:film-length-avg} for $\theta = 0$, since the latter simulations include some bubble attraction due to our finite film thickness (assumed to be zero in the equation's derivation). From this figure, the film lengths decrease initially with liquid fraction, as the bubbles become less deformed. For repulsive $\Pi_\text{D}$, the lengths vanish as the bubbles separate at the jamming transition, while a plateau is seen for $\theta \geq 0$ due to flocculation. Our results for $\theta > 0$ are in qualitative agreement with \citeauthor{2021-feng} \cite{2021-feng}, who measured the total length of all films in foams simulated using a different model, and are consistent with the interpretation that $\theta > 0$ caused films to be retained at $\phi > \phi_\text{c}$ in the ISS coarsening experiments \cite{2023-pasquet-b}.
	
	We note that Eq.~\eqref{eqn:film-length-avg} predicts that the plateau satisfies $\langle L / P \rangle \approx \langle n \rangle \, \theta / \pi$, since $\Pi_\text{O} \approx 0$ in a flocculated foam \cite{1983-princen}. We note that similar predictions of the mean film area in flocculated three-dimensional foams have been made (which instead vary as $\theta^2$) \cite{2023-durian-pre,2023-pasquet-b}.
	
	The error in Eq.~\eqref{eqn:film-length-avg} is larger for smaller $\phi$, likely due to the  correlations between neighbouring bubbles mentioned in Sec.~\ref{sub:coa:bubbles}. We note that the exact mean $\langle L / P \rangle$ predicted by Eq.~\eqref{eqn:film-length-approx} behaves similarly, although is a slightly worse approximation to the data in the plateau.
	
	\subsubsection{Growth rates \label{ssb:coa:growth-stats}}
	
	Next, we consider the averaged bubble growth rates. Noting that $\langle \dot{A} \rangle = 0$ because the system is closed and the gas is taken as incompressible, we consider the root-mean-square growth rate $\sqrt{\langle \dot{A}^2\rangle}$. Since this measures typical instantaneous bubble growth rates (in absolute value), we use it as a proxy for the initial rate at which the foam would coarsen with time.
	
\begin{figure}
		\includegraphics[width=1.0\columnwidth]{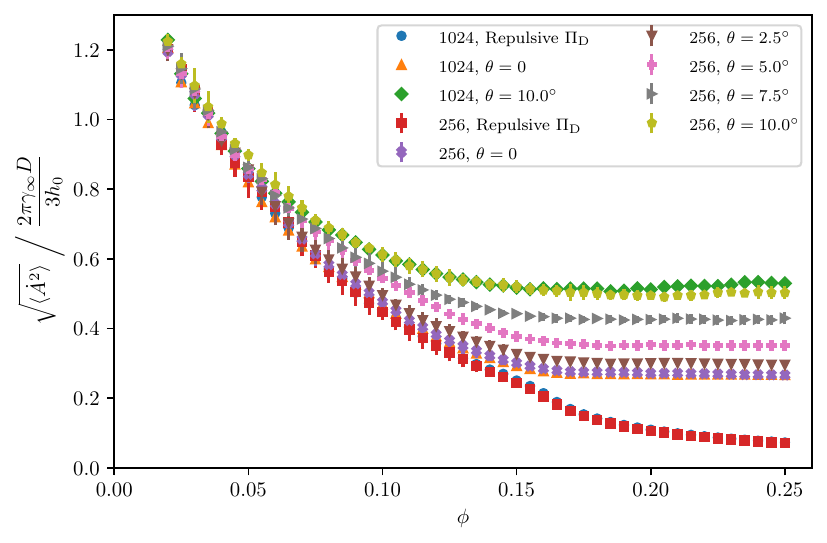}
		\caption{Root-mean-square growth rate versus $\phi$, for repulsive $\Pi_\text{D}$ and various $\theta$. Results for single $1024$-bubble runs are shown, along with the mean of $5$ runs with $256$ bubbles (the error bars give the sample standard deviation). \label{fig:rms-growth}}
	\end{figure}

	In Fig.~\ref{fig:rms-growth}, we plot the variation of $\sqrt{\langle \dot{A}^2\rangle}$ with $\phi$, for several values of $\theta$. As expected, the coarsening rate decreases with increasing $\phi$ for small liquid fractions, since the Plateau borders grow larger and further frustrate the flow of gas \cite{2013-roth}. The contact angle does not have a strong effect in this regime.
	
	For repulsive $\Pi_\text{D}$, the coarsening rate decreases for all $\phi$, since the gas flow between separated bubbles decreases as their separation increases \cite{2017-schimming} (see Fig.~\ref{fig:foam-struct-25p}). However, as observed for the film lengths, the presence of bubble attraction induces a plateau in the coarsening rate. As noted in Sec.~\ref{sub:coa:equilibrium}, the bubbles flocculate~\cite{2018-cox}, thereby retaining films that allow for efficient gas transfer. The typical film length increases with $\theta$, and so the plateau values likewise increase. We recall that there is bubble attraction for $\theta = 0$ in our model, due to the finite film thickness. Hence, we are consistent with reports that flocculation in emulsions increases the coarsening rate \cite{2008-djerdjev}, along with the interpretation that $\theta > 0$ caused the $\langle R \rangle$ growth exponent (see Sec.~\ref{sec:intro}) to remain equal to $1/2$ (i.e.\ its value for dry foams) at higher $\phi$ than expected in experiments on the ISS \cite{2023-pasquet-b}. However, obtaining growth exponents from our simulations would require following coarsening through time, which is not feasible for large enough systems at present.
	
	For similar reasons, we have not investigated the relationship between $\sqrt{\langle\dot{A}^2\rangle}$ and the initial coarsening rate, measured for example by $d \langle A \rangle / d t$ (with system size large enough, and time resolution sufficiently coarse, that $\langle A \rangle$ is smooth \cite[p.\ 105]{2013-cantat}). In particular, it has been suggested that the structure of flocculated foams may change substantially with time under coarsening \cite{2023-pasquet-b}. Furthermore, by comparing with simulations using bubble areas directly from a Voronoi tessellation, we believe that the variation shown in Fig.~\ref{fig:rms-growth} depends quantitatively on polydispersity, but not qualitatively.
	
	Finally, we recall that a bubble's border-blocking growth rate is proportional to the pressure difference between a bubble and its neighbour, multiplied by the ratio of film length to thickness, and summed over each film \cite{2013-cantat,1991-bolton}. Therefore, Eqs.~\eqref{eqn:eff-yl} and~\eqref{eqn:film-length-approx} can be combined to give an approximate border-blocking bubble growth law
	\begin{align}
		\dot{A}_\text{BB} &\approx \frac{4 \pi \gamma_\infty D}{h_0} \left(\frac{R}{R_\text{C}} - 1\right) 
			\frac{\bar{\Pi} \bar{R} + n \, \theta / \pi}{1 + (2 \bar{\Pi} + 1) \bar{R}} ,
			\label{eqn:growth-rate-lemlich-type}
	\end{align}
	where the foam's critical radius $R_\text{C}$ (i.e.\ the radius of the neighbouring bubbles, taken as constant for the foam) is determined by $\langle \dot{A} \rangle = 0$~\cite{1978-lemlich,2023-pasquet-b}. Again, we assume $\theta \lesssim 10^\circ$. This gives the growth rate solely in terms of the bubble radius and foam properties (notwithstanding the aforementioned need to relate $n$ to $R$), and thus is a (two-dimensional) generalization of \citeauthor{1978-lemlich}'s approach \cite{1978-lemlich} to allow for a variation in film length with bubble size. Recalling Fig.~\ref{fig:film-length-v-radius}, smaller bubbles have shorter films relative to their size, and thus experience a larger degree of border blocking \cite{1991-bolton}. While Eq.~\eqref{eqn:growth-rate-lemlich-type} does not accurately give the growth rates of individual bubbles, it may be fruitful to analyse the size distributions it predicts in the coarsening scaling state~\cite{1978-lemlich,1997-yarranton}, particularly whether it can reproduce the large population of small bubbles observed recently by \citeauthor{2023-galvani} \cite{2023-galvani} through the enhanced border blocking of such bubbles.
	
	In proposing Eq.~\eqref{eqn:growth-rate-lemlich-type}, we follow the suggestion of \citeauthor{2023-pasquet-b} \cite{2023-pasquet-b} to investigate growth laws which generalize \citeauthor{1978-lemlich}'s approach \cite{1978-lemlich} by accounting for correlations between bubble pressures and film lengths. A three-dimensional growth law of a different form, derived by considering gas flow through the bulk liquid (rather than the films), but which also incorporates correlations between pressure and efficiency of diffusion, was studied numerically by \citeauthor{1997-yarranton} \cite{1997-yarranton}.

	\section{Conclusions \label{sec:conclusions}}
	
	We have described simulations for studying coarsening in two-dimensional wet aqueous foams, with a structural model based upon the work of \citeauthor{2014-kahara} \cite{2014-kahara} and \citeauthor{2018-boromand} \cite{2018-boromand}, and a coarsening approximation inspired by the analytical work of \citeauthor{2008-marchalot} \cite{2008-marchalot} and \citeauthor{2017-schimming} \cite{2017-schimming}. Our methods allow for  arbitrary liquid fractions $\phi$, and support contact angles $\theta$ up to about $10^\circ$. The interfaces interact through a model disjoining pressure, although our selected form could be swapped for another.
	
	We have applied this model to polydisperse foams, containing $256$ to $1024$ bubbles. $\phi$ has been varied by gradually increasing its value within particular foam samples, following the approach of \citeauthor{1990-bolton} \cite{1990-bolton} and \citeauthor{2018-cox} \cite{2018-cox} using a different model. We have analysed the coarsening-related properties of these foams. For $\theta \lesssim 10^\circ$, the bubble pressures $p$ were found to obey an effective Young-Laplace law relating them to the osmotic pressure $\Pi_\text{O}$, and we derived an approximation for the bubble film lengths $L$. When its prediction of the averaged film length is estimated, it takes an equivalent form to a previously-proposed approximation in three dimensions for $\theta = 0$ \cite{2021-hohler,2023-pasquet-b}, and also agrees well with our simulations.
	
	Turning to the coarsening-induced bubble growth rates, we showed that their root-mean-square values plateau with increasing $\phi$ in foams with attractive bubble interactions, due to the previously-simulated \cite{2018-cox} flocculation of bubbles. The effective neighbour number $n_\text{eff}$ of \citeauthor{2012-fortuna} \cite{2012-fortuna} was found to be closely related to the growth rate (determining whether bubbles grow or shrink to a good approximation) whenever bubbles had films between them, caused by either the osmotic pressure or flocculation.
	
	We also compared our simulations of gas flow through Plateau borders with existing predictions \cite{2017-schimming}. Qualitative agreement was found for bubbles at small $\phi$, along with quantitative agreement at the jamming transition for zero bubble attraction. An underestimate of the border gas flow rates at small $\phi$ is believed to arise from ambiguity in the film lengths due to our finite film thickness.
	
	Our simulations remain computationally expensive, and so we have not considered the time-dependence of foams under coarsening here. However, it may be possible to obtain the timescale over which coarsening occurs in small systems, and to observe its relationship to the root-mean-square growth rates we considered. Such timescales have been investigated by \citeauthor{2018-khakalo} \cite{2018-khakalo}, without bubble attraction.
	
	The approach we use seems most suited to accurately simulating foams on a small scale (with around $1000$ bubbles), in order to develop models of their properties. These could then be implemented in large-scale coarsening simulations using the bubble model \cite{1995-durian}, for example. The analytical approximations we have presented here may be a starting point for this process, recalling that our expressions for the bubble pressures and film lengths may be combined to give an approximate growth law. As noted above, the properties of the resulting scaling states could be analysed \cite{1978-lemlich,1997-yarranton}, which may contribute to the interpretation of experiments of coarsening in wet foams performed in microgravity \cite{2021-born,2023-galvani,2023-pasquet-b}.
	
	As noted in Sec.~\ref{sec:methods}, the implementation of our simulation model could be improved, including the equilibration algorithm, which may result in larger foams becoming feasible. The coarsening approximation could also be tested against accurate solutions of Laplace's equation within the foam's liquid, partly to gauge the importance of long-range gas transfer \cite{2022-hohler} for which the approximation is expected to be poor. Our large equilibrium film thickness $h_0$ could be varied to determine its effect on our results (we expect it to have increased the proportion of gas transfer through Plateau borders \cite{2017-schimming}), along with our small derivative of the disjoining pressure at $h_0$.
	
	Our analytical approximations could be extended by investigating the relationship between bubble radius and neighbour number, and between the osmotic pressure and the mean neighbour number. By using an independent empirical expression for the osmotic pressure, similar to those described by \citeauthor{2021-hohler} \cite{2021-hohler}, these approximations could be made fully predictive. Finally, as noted by \citeauthor{2023-pasquet-b} \cite{2023-pasquet-b}, coarsening in flocculated foams and emulsions at large $\phi$ has not been widely studied. We note that the structure of flocculated systems differs greatly as the contact angle is varied \cite{1993-bibette}.
	
	Another extension is to perform similar studies of three-dimensional wet foams (albeit for fewer bubbles). We have adapted the simulation model to this case, noting that the approach of \citeauthor{2018-boromand} \cite{2018-boromand} has been similarly adapted by \citeauthor{2021-wang} \cite{2021-wang}. A comparable three-dimensional approach for biological cells was also used by \citeauthor{2020-van-liedekerke} \cite{2020-van-liedekerke}. Such models have not yet been applied to the simulation of foams, to our knowledge, although simulations with zero film thickness have been performed by \citeauthor{2020-kraynik} \cite{2020-kraynik}.
	
	The data upon which this work is based is available at Ref.~\cite{2023-morgan-data}.
	
	\begin{acknowledgments}
		The authors wish to acknowledge useful discussions with Tudur Davies, Fran\c{c}ois Graner, and Reinhard H\"ohler, along with Kenneth Brakke for developing the {\sc Surface Evolver}. JM acknowledges funding from ESA REFOAM MAP contract 4000129502.
	\end{acknowledgments}
	
	\appendix*
	
	\section{Analytical model for film lengths \label{app:analytical}}
	
	We now derive the approximation for a bubble's film length $L$ used in Sec.~\ref{sec:coarsen-poly}, assuming mechanical equilibrium (as holds in our simulations). The approach was inspired by \citeauthor{2021-hohler} \cite{2021-hohler}, and is based upon a standard equilibrium result relating the stress tensor $\bar{\bm{\tau}}$, averaged (in the manner of \citeauthor{1970-batchelor} \cite{1970-batchelor}) over a domain $\mathcal{S}$ of area $A$, to the force per unit length $\mathbf{f}$ applied at its boundary $\Gamma$ \cite[pp.\ 6--7]{1986-landau}:
	\begin{align}
		\bar{\tau}_{i j} = \frac{1}{2 A} \oint_\Gamma d l \, (f_i r_j + f_j r_i) ,
			\label{eqn:int-ext-stress-rel}
	\end{align}
	where $\mathbf{r}$ is the displacement of a point on $\Gamma$ from an arbitrary fixed origin. Let $\mathcal{S}$ be the domain of a bubble (of effective radius $R$), including its interface (which is then equivalent to $\Gamma$). Our definition $\bar{p} = -\tfrac{1}{2} \Tr \bar{\bm{\tau}}$ (Sec.~\ref{sub:coa:bubbles}) then gives
	\begin{align}
		\bar{p} = -\frac{1}{2 \pi R^2} \, \oint_\Gamma d l \ \mathbf{f} \cdot \mathbf{r} , 	
			\label{eqn:avg-press-bound-rel}
	\end{align}
	This relates the reduced pressure over the bubble to the forces applied at its interface. Let $\mathbf{n}$ be the outward unit normal to $\Gamma$, and $\Pi_\text{D}$ the corresponding disjoining pressure. Then $\mathbf{f} = -\Pi_\text{D} \, \mathbf{n}$ since we set the liquid pressure to zero. Let us take the limit of small film thickness. Therefore, where $\Gamma$ adjoins Plateau borders, $\mathbf{f} = \bm{0}$. On the interface of a film shared with a bubble of pressure $p_i$, we have $\Pi_\text{D} = (p + p_i) / 2$ \cite[p.\ 90]{1998-exerowa}. Under a mean-field \cite{1978-lemlich} approximation, let $p_i \approx \Pi_\text{C}$ \{given in turn by Eq.~\eqref{eqn:cap-osm-rel} \cite{2021-hohler}, again dropping the factor $1/(1-\phi)$\}, and use Eq.~\eqref{eqn:eff-yl} for $p$. Thus, adjoining the film interfaces,
	\begin{align}
		\Pi_\text{D} \approx \frac{\gamma_\infty}{2 R} \left[1 + (2 \bar{\Pi} + 1) \bar{R}\right] ,
			\label{eqn:dis-press-film}
	\end{align}
	where $\bar{\Pi} = \Pi_\text{O} / (\gamma_\infty / R_{2 1})$ and $\bar{R} = R / R_{2 1}$.
	
	At the points of $\Gamma$ where the films and Plateau borders meet, a transversal line tension acts to maintain equilibrium for $\theta > 0$ \cite{1985-kralchevsky}. This is exerted by the medium-range attraction in $\Pi_\text{D}$ (see Fig.~\ref{fig:dis-press}). We model this interaction as a singular contribution to $\mathbf{f}$; i.e.\ as forces $\mathbf{F} = \gamma_\infty \, \mathbf{n} \sin\theta \approx \gamma_\infty \theta \, \mathbf{n}$ \cite{1985-kralchevsky} applied at the film-border transition points (we again consider $\theta \lesssim 10^\circ$).
	
	Substituting the above expressions for $\mathbf{f} = -\Pi_\text{D} \, \mathbf{n}$ into Eq.~\eqref{eqn:avg-press-bound-rel}, and making the rough approximation $\mathbf{r} = R \, \mathbf{n}$ (i.e.\ that the bubble is circular), we obtain
	\begin{align}
		\bar{p} \approx \frac{\gamma_\infty}{2 \pi R} \left\{\left[1 + (2 \bar{\Pi} + 1) \bar{R}\right] \frac{L}{2 R} - 2 n \theta\right\}, \label{eqn:avg-press-approx}
	\end{align}
	where $n$ is the bubble's number of neighbours. Finally, we use the approximation $\bar{p} \approx \Pi_\text{O}$ discussed in Sec.~\ref{sub:coa:bubbles}. Rearranging Eq.~\eqref{eqn:avg-press-approx}, and using that the perimeter $P \approx 2 \pi R$ \cite{2012-kraynik}, then gives our approximation for the bubble's film length $L$, Eq.~\eqref{eqn:film-length-approx}.

\end{document}